# 2022
# Superfacility Project Report

Lawrence Berkeley National Laboratory
## Computing Sciences

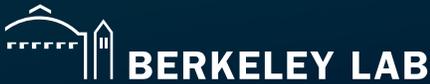
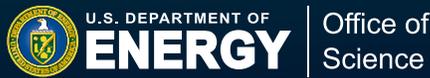

Lawrence Berkeley National Laboratory  |  1 Cyclotron Road  |  Berkeley, CA94720-8148

# LBNL Superfacility Project Report

Debbie Bard, Cory Snavely, Lisa Gerhardt, Jason Lee, Becci Totzke, Katie Antypas, William Arndt, Johannes Blaschke, Suren Byna, Ravi Cheema, Shreyas Cholia, Mark Day, Bjoern Enders, Aditi Gaur, Annette Greiner, Taylor Groves, Mariam Kiran, Quincey Koziol, Tom Lehman, Kelly Rowland, Chris Samuel, Ashwin Selvarajan, Alex Sim, David Skinner, Laurie Stephey, Rollin Thomas, Gabor Torok
*Lawrence Berkeley National Laboratory, Computing Sciences Area, Berkeley, California 94720*

# Executive Summary

The Superfacility model is designed to leverage HPC for experimental science. It is more than simply a model of connected experiment, network, and HPC facilities; it encompasses the full ecosystem of infrastructure, software, tools, and expertise needed to make connected facilities easy to use. The three-year Lawrence Berkeley National Laboratory (LBNL) Superfacility project was initiated in 2019 to coordinate work being performed at LBNL to support this model, and to provide a coherent and comprehensive set of science requirements to drive existing and new work. A key component of the project was the in-depth engagements with eight science teams that represent challenging use cases across the DOE Office of Science.

By the close of the project, we met our project goal by enabling five of our science application engagements to demonstrate automated pipelines that analyze data from remote facilities at large scale, without routine human intervention. In several cases, we have gone beyond demonstrations and can now provide production-level services for their experiment teams:
- Dark Energy Spectroscopic Instrument (DESI): Automated nightly data movement from telescope to NERSC and deadline-driven data analysis with telescope operations starting in 2020.
- Linac Coherent Light Source (LCLS): Automated data movement and analysis from several experiments running at high datarate end stations during 2020 and 2021
- Lux-Zeplin experiment (LZ): Automated 24/7 data analysis from the dark matter detector, with commissioning starting in June 2020
- National Center for Electron Microscopy (NCEM): Automated workflow pulling data from the 4D STEM camera to Cori for near-real-time data processing, starting in late 2021.

Other science partners from the Advanced Light Source (ALS), the Rubin Observatory Dark Energy Science Collaboration (DESC), the Joint Genome Institute (JGI) and the KSTAR fusion science team also made significant advances in their workflows. To achieve this goal, the Superfacility project team developed tools, infrastructure, and policies for near-real-time computing support, dynamic high-performance networking, data management and movement tools, API-driven automation, HPC-scale notebooks via Jupyter, authentication using Federated Identity and container-based edge services supported via Spin. The Superfacility project team included members from the National Energy Research Scientific Computing center (NERSC), the Energy Sciences network (ESnet) and the Computer Science (CS) research divisions at LBNL. The lessons we learned during this project provide a valuable model for future large, complex, cross-disciplinary collaborations.

Superfacility work continues at LBNL, supporting the tools and infrastructure we developed during the project and initiating new work in response to emerging science needs. There is a pressing need for a coherent computing infrastructure across national facilities, and LBNL's Superfacility project is a unique model for success in tackling the challenges that will be faced in hardware, software, policies, and services across multiple science domains.

# 1: Overview

Technology is transforming science in many ways. New instruments and higher fidelity detectors are producing an explosion in data rates. Scientists are seeking new computing hardware, software, algorithms and frameworks to process and analyze this data. The net result is a massive potential for scientific insight and discovery in fields ranging from genomics to cosmology to materials science. Increasingly, scientists are turning to HPC to provide the computing power necessary to manage and analyze their data, which is "big" both in terms of quantity and also in complexity. However, HPC systems have not historically been designed to meet these needs, so developing workflows to analyze experimental data on HPC is labor intensive and painful. Scientists have needed to work around existing policy and technology gaps, which can significantly hinder scientific progress.

If we remove these hurdles, we can open up new fields with tremendous potential for breakthrough. For example, HPC-scale real-time automated analysis of data streaming from a running experiment can give scientists fast feedback on the progress of their experiment and save valuable instrument time. Large-scale simulations coupled with near-real-time data analysis can guide experiments in a digital twin model, and automated data movement, management and job control can release valuable scientist time to focus on interpretation of results. The Superfacility project at LBNL was designed to provide scientists with the high-performing, easy-to-use, scalable, and automated tools to run their workflows across geographically distributed experiment, network and compute facilities, with minimal human intervention.

The Superfacility concept was first proposed at LBNL almost ten years ago as a model to leverage HPC for experimental science. In its simplest form, it can be considered a model of connected facilities – an experiment facility connected to a high performance network (usually ESnet) connected to a supercomputing center. But to make this a productive environment for a scientist, we also need to consider the full ecosystem of infrastructure, software, tools, and expertise needed to make connected facilities easy to use.

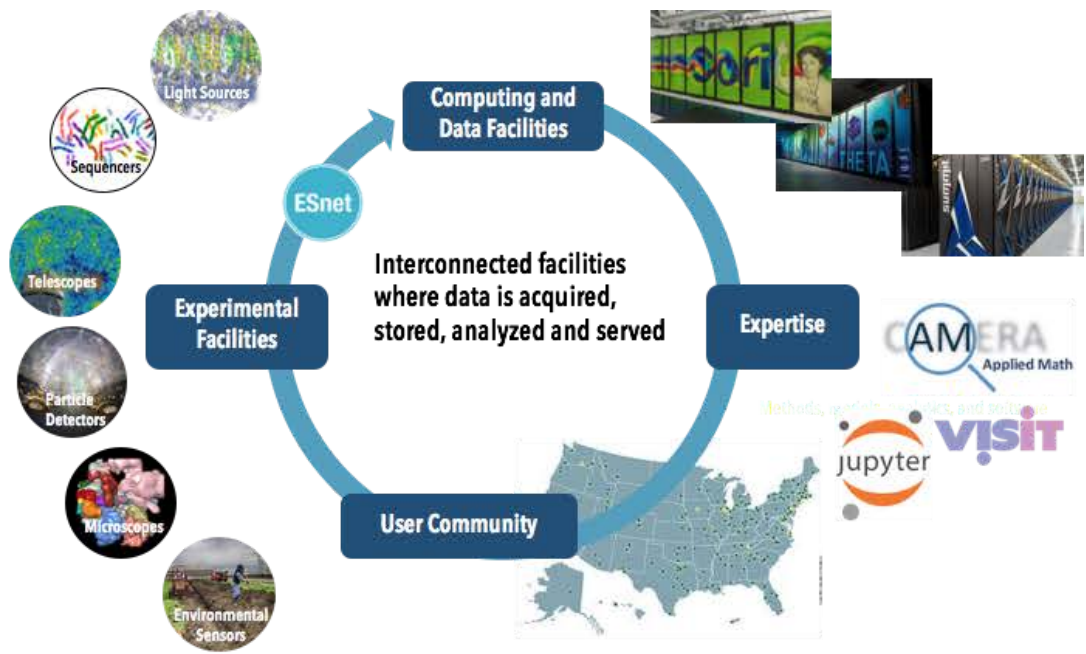

*Figure 1. The Superfacility Model.*

In the 2019 LBNL Computing Sciences Strategic Plan,[1] the Superfacility model was identified as a key goal for the CS Area. Several areas of work were described, including user engagement, data management, automation and edge computing. Work had been ongoing in these areas for some years, but it tended to be done via one-off projects or demos with specific science teams. The work was isolated, and had not coalesced into something that could be widely used by multiple science teams. The Superfacility project was created to coordinate these efforts, provide a coherent set of requirements, and ensure that the main goals laid out in the 2019 strategic plan could be met. The project is a key strategic priority for NERSC and ESnet and work was conducted as part of the facilities' base funding profile. The project did not have additional funding attached, nor did it aim to cover all superfacility-related work being conducted at LBNL (see Section A.4 for details). This gave a useful boundary to the project scope: we aimed to coordinate existing and planned work to ensure the greatest possible impact, and to initiate new work that could be achieved in the given time frame.

We adopted a requirements-driven process: via multiple iterations with our selected science engagements (see Section 2 for details on how we selected these teams) we were able to set an initial scope and prioritization of work to ensure that we focussed on the areas that were most important to our science teams. This may not have been the work that we had initially thought of as most important; for example, support for resilience workflows emerged as a key requirement for several science teams that was not in our initial plan. Science moves quickly, and by conducting regular directed requirements conversations with our engagements (for example, via our Year-2 survey), as well as the day-to-day interactions our engagement leads had with the science teams, we were able to update our requirements to make sure our work stayed relevant.

---

[1] https://cs.lbl.gov/assets/Uploads/19-CS-5660-Computing-Sciences-StratPlan-043019-Online.pdf

The project goal was designed to be achievable in three years, and at the same time to be ambitious enough to make a substantive change in the way science teams use ESnet and NERSC. We identified the capabilities required based on our requirements conversations with science teams.

> **Project Goal:**
> By the end of CY 2021, 3 (or more) of our 7 science application engagements will demonstrate automated pipelines that analyze data from remote facilities at large scale, without routine human intervention, using these capabilities:
> - Near real-time computing support
> - Dynamic, high-performance networking
> - Data management and movement tools, incl. Globus
> - API-driven automation
> - HPC-scale notebooks via Jupyter
> - Authentication using Federated Identity
> - Container-based edge services supported via Spin

In order to enable the Superfacility model of science and to support increasing data rates and complex workflows, we need to focus on automated pipelines. Automation is a key part of the project goal for several reasons. We aim to reduce the number of humans in the loop that are generally required to run this kind of complex workflow, which enables users to scale their science, analyzing large-scale data or receiving results faster. It also enables NERSC to support many more experimental data analysis projects. For example, prior to the Superfacility project, a NERSC staff member was required to be present to supervise data movement and analysis for experiments at the Linac Coherent Lightsource (LCLS). Automating these processes is a key part of our strategy. We see an increasing demand for HPC-scale data analysis from experiment user facilities, which carries with it a corresponding increase in people using NERSC. Without automation, we will not be able to scale user support to these new communities. Automation also allows us to give these user facilities more autonomy in how they handle their own users (such as managing their allocations and users at NERSC via the [PI Dashboard](#)) without needing to ask NERSC to act on their behalf.

This report describes the science requirements driving our work in [Section 2](#), and the technical work we undertook to support that science in [Section 3](#). A summary of the impact of this work on the science engagement is given in [Section 4](#). We describe the lessons we learned during this project in [Section 5](#), which we hope will be of use to others planning any cross-disciplinary project. The work achieved in this project is substantial, but there is much still to do; some short- and medium-term priorities for future work are given in [Section 6](#). We also describe how we constructed and managed the project in [Appendix A](#).

# 2: Science

A key principle behind our work in the Superfacility project was to ensure that our work scales across multiple science domains of interest to the DOE Office of Science. We wanted to avoid unsustainable one-off solutions that place large support burdens on both science teams and facility staff, and instead design an ecosystem of interconnected technologies that would fulfill the needs of many of our users. To this end, we selected a set of seven (later eight) science engagements that would drive our requirements and provide us with highly engaged beta testers. We chose these engagements carefully, based on:
- Science area: representing experiment teams across the DOE SC
- Workflow complexity: representing a range of complicated end-to-end workflows that have data sources within LBNL, elsewhere in the U.S., and internationally.
- Scale: from relatively small teams (a few scientists) to the largest teams at NERSC (a few hundred scientists); with compute needs ranging from a few million to a few hundred million hours a year; and with data rates in the TB/year to PB/year.
- Timeframe: from experiments taking data right now to more forward-looking projects preparing for future large data rates.

Our science engagements provide us with use cases that match a large range of user requirements – not just in the experimental science space, but that also extend to the full NERSC user base.

| Science team | Science area | Data source | Scale (2021) | Timeframe |
|---|---|---|---|---|
| Advanced Light Source (ALS) | BES Lightsource | DOE user facility at LBNL | 100s of users, 50M NERSC hours/year, 600TB/year, (10Gb/sec streaming) | Operating throughout project. Upgrade in 2025. |
| Rubin Observatory Dark Energy Science Collaboration (DESC) | HEP Optical survey telescope | Telescope in Chile (2024) Simulated data at NERSC, ALCF and IN2P3 (France) | 100s of users, 150M NERSC hours/year, 2 PB/year | Telescope operations start in 2024. Simulated data produced throughout project. |
| Dark Energy Spectroscopic Instrument (DESI) | HEP Spectroscopic survey telescope | Telescope in Arizona | 100s of users, 200M NERSC hours/year, 500TB/year (10GB/night) | Nightly data taking started in 2020. |
| Joint Genome Institute (JGI) | BER Genomics | DOE user facility at LBNL | 100s of users, 75M NERSC | Operating ~continuously |

|  |  |  | hours/year, 500TB/year | throughout project. |
|---|---|---|---|---|
| Korea Superconducting Tokamak Advanced Research (KSTAR) | FS Tokamak | Fusion facility in South Korea | 10s of users, 145M NERSC hours/year, 20TB / year, (~10GB/hour streaming) | Will operate during KSTAR run campaigns, 1-2 per year. |
| Linac Coherent Lightsource (LCLS) | BES Lightsource | DOE user facility at SLAC | 100s of users, 12M NERSC hours/year, 1PB /year (100 Gb/sec streaming, bursty) | Operating now. Using NERSC for specific ~bimonthly experiments |
| LUX-ZEPLIN (LZ) | HEP Dark Matter | Experiment in Sanford Underground Research Facility | 100s of users, 20M NERSC hours/year, 1PB/year (~GB/hour streaming) | 24/7 data taking started in 2021. |
| National Center for Electron Microscopy (NCEM) | BES Electron Microscope | DOE User Facility at LBNL, high frame-rate microscope | 10s of users, 1M NERSC hours/year, 600TB/year (100Gb/sec streaming) | Operating throughout project. Started using NERSC for running experiments in 2021. |

## Engagement model

A key part of the success of this project was assigning dedicated science liaisons to each of our science engagements. This is a single person who was the expert for questions about that science team and who worked closely with them (often this was a science team they were already working closely with). Having NERSC experts on the science needs allowed us to streamline a lot of the requirements process and gave the science teams a single point of contact for problems and questions. This support approach is not scalable to the whole NERSC user base but served a very valuable purpose in this project.

Our engagement model was designed to be as efficient as possible. We wanted to limit the number of requests we made to the science teams, and at the same time ensure we heard everything they needed to tell us. To this end we had four stages of engagement:
1. Pre-project: Initial requirements conversations with the science engagement leads, followed by discussion with the full Superfacility team. These were fairly unstructured and were designed to let the science teams tell us what they needed in their own words.

2. Fall 2019: Requirements survey, comprising a single survey with questions from most of the technical work areas about specific features or capabilities that were being considered.
3. Sporadic: When a new capability was close to deployment, science teams who had expressed a particular interest in that capability were invited to be beta-testers by the developers. This ensured the tool could be adjusted as necessary and new features identified.
4. Continuous: science engagement leads had regular conversations with their science teams, often in the form of bi-weekly or monthly meetings. These close engagements will continue after the end of this project.

The following sections introduce the science teams and discuss their requirements. We describe how the Superfacility project has helped them achieve their science goals in Section 4.

## 2.1: ALS

> **Key Superfacility needs:** NESAP, Policies, Jupyter, Scheduling, Resiliency, Federated ID, API, Spin, Self-managed Systems, Data movement, Data management.

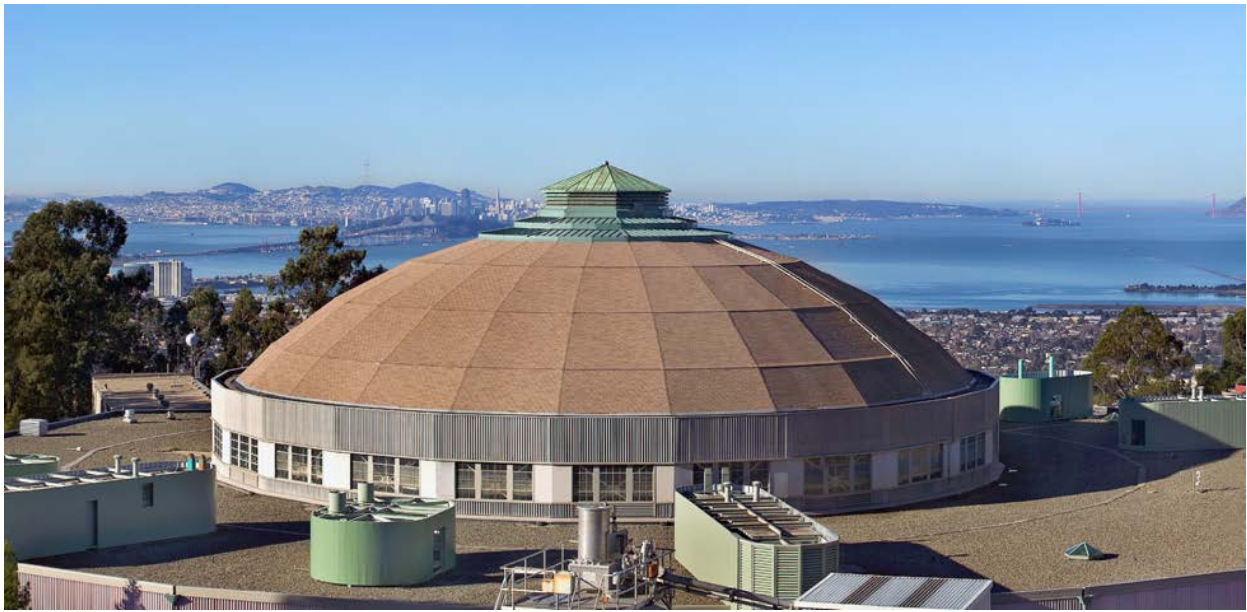

*Figure 2. The ALS building at LBNL.*

The Advanced Light Source[2] (ALS), a synchrotron radiation facility situated at LBNL, is one of DOE's five large light sources. It comprises about 40 beamlines with numerous experimental endstations, where scientists from around the world ("users") can conduct research in a wide variety of fields, including materials science, biology, chemistry, physics, and the environmental sciences. ALS serves a user community of roughly 2,000 users per year. Like other light sources, it faces the challenge[3] that current and future upgrades to its storage rings will vastly increase the amount of data that is generated. This flood of data will make local data storage and computing unfeasible in the near future.

---
[2] https://als.lbl.gov/

ALS became partner in the Superfacility project to address this challenge, with a focus on:
- GPU-enabled analysis code via NESAP
- Modernizing data management, movement, access and archiving, including use of Spin and Federated ID
- Using HPC for near-real-time feedback for their experiments, including interactive data analysis via Jupyter and resilience to operate when NERSC is unavailable
- Empowering users to independently analyze their data even after their experiments are over (hand off).

## 2.2: DESC

> **Key Superfacility needs:** NESAP, Policies, Jupyter, Resiliency, Federated ID, API, Spin, SDN, Data movement, Data management.

The Dark Energy Science Collaboration[4] (DESC) consists of cosmologists from around the world who will use data from the Rubin Observatory to understand the nature of Dark Energy. The Rubin Observatory is scheduled to begin operations in 2024, so the DESC is currently focused on developing analysis pipelines that will extract science from the vast datasets produced by Rubin. These pipelines are being refined using data challenges, which are end-to-end cosmology-to-pixels-to-catalog simulations, using precursor datasets from existing observatories, and eventually using data from the commissioning phase of the Rubin Observatory.

NERSC is the main computing site for DESC, and we have worked closely with the DESC computing team to support their large-scale simulation pipelines, as well as their use of Jupyter for collaboration-wide data analysis.

---

[3] https://physicstoday.scitation.org/do/10.1063/PT.6.2.20200925a/full/
[4] https://lsstdesc.org/

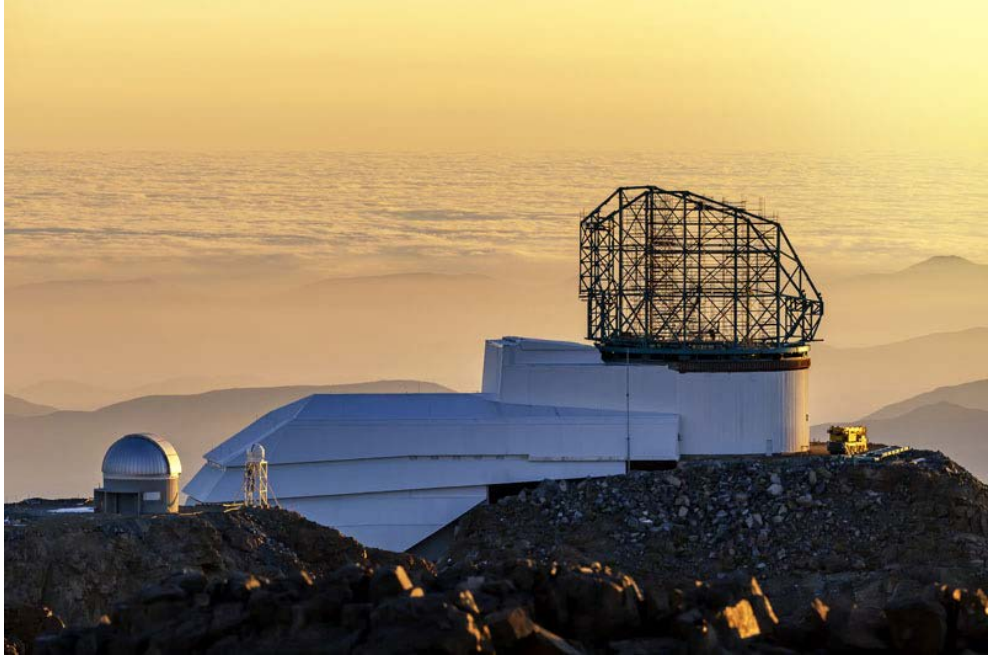

*Figure 3. The Rubin Observatory in Chile, under construction in 2019.*

DESC partnered with the Superfacility project to support their current simulation campaign and their future data analysis. Their requirements included:
- Optimizing simulation code for future architectures via the NERSC Exascale Science Applications Program (NESAP)
- Coordinating long-running simulation and data processing campaigns and managing the resulting massive datasets, using Spin. Data is also shared publicly via a data portal hosted on Spin.
- Interactive analysis of simulated data via Jupyter
- Management of a collaboration of hundreds of NERSC users.

## 2.3: DESI

**Key Superfacility needs:** NESAP, Policies, Jupyter, Scheduling, Resiliency, Spin, Self-managed Systems, Data movement, Data management.

DESI[5] is the Dark Energy Spectroscopic Instrument, a cosmology experiment designed to generate the most detailed 3D map to date of the universe. Using the DESI spectroscopic instrument installed in the Mayall telescope at Kitt Peak, AZ, this survey will be conducted over 14,000 square degrees of the sky in the Northern Hemisphere over five years. This map will be used to better understand the physics of dark energy and its role in the expansion history of the universe.

---
[5] https://www.desi.lbl.gov/

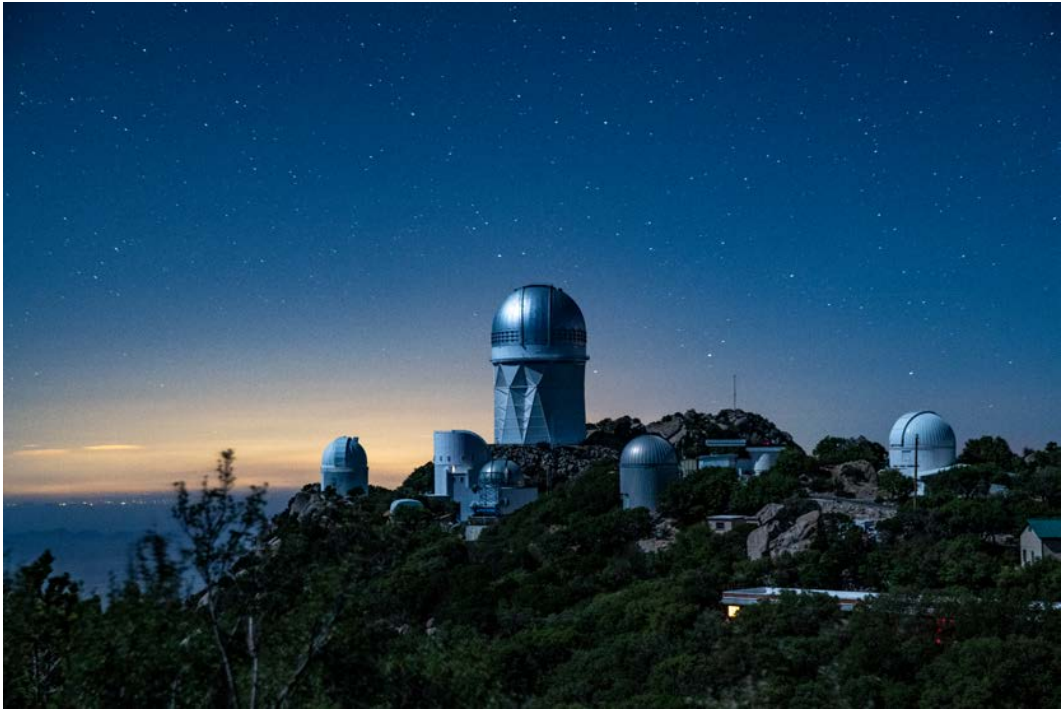
*Figure 4. The Mayall Telescope at Kitt Peak National Observatory.*

DESI has 5,000 robotically controlled fiber positioners, each of which can image a separate object, such as a galaxy or a quasar, during an exposure. Thirty CCD spectrograph images with raw spectral data are generated per exposure. DESI can generate exposures every 10 minutes and can complete up to 50 exposures per night.

DESI uses NERSC to quickly process DESI's raw exposure data into usable scientific results that include the calculation of redshifts, which can in turn be used to determine the location of the object. Being able to process the nightly data quickly (the common phrase is "redshifts by breakfast") is important to DESI because it allows them to quickly assess the quality of the exposures and plan their next night of observing accordingly.

DESI is a long-term user of NERSC for both simulation production and developing analysis pipelines. DESI partnered with the Superfacility project to support the following requirements:
- GPU-enabled analysis code via the NESAP program
- Operations database mirror, QA monitoring, and data access via Spin
- Deadline-driven computing for nightly data analysis and resiliency to NERSC outages
- Interactive data analysis using Jupyter.

## 2.4: JGI

> **Key Superfacility needs:** Policies, Resiliency, Federated ID, API, Spin, Data movement, Data management.

The role of the [Joint Genome Institute](https://jgi.doe.gov/)[6] (JGI) is to advance genomics understanding in support of DOE missions related to clean energy generation, environmental characterization, and pollution remediation. The JGI is a DOE user facility that employs hundreds of staff and serves 2,000 project users and more than 10,000 data users. Services provided to those users include high-throughput sequencing, DNA design and synthesis, metabolomics, and computational analysis.

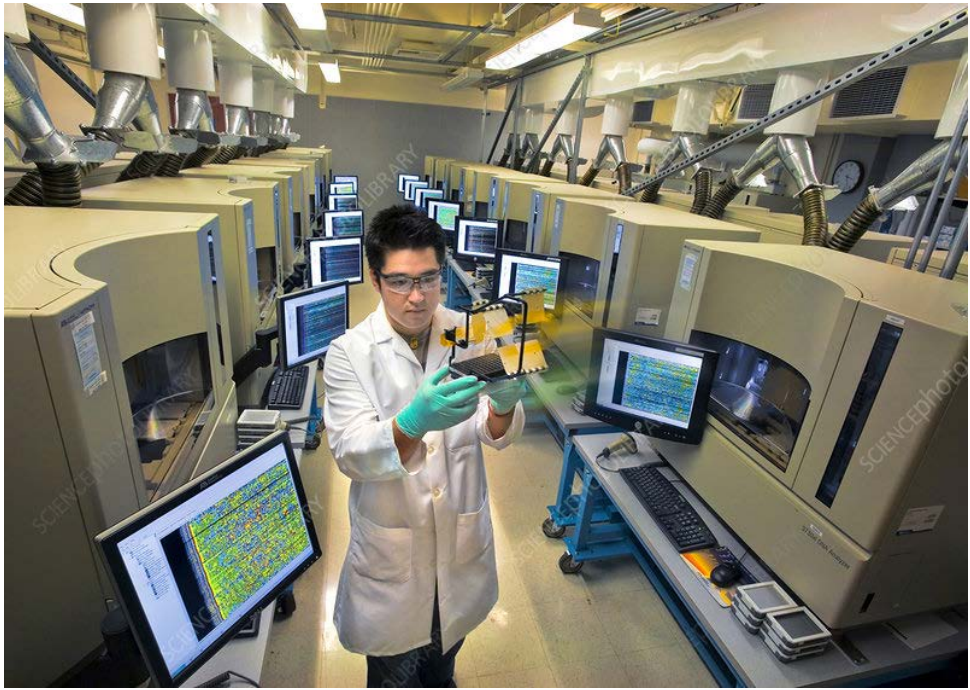

*Figure 5. Sequencers at the JGI.*

An important distinction between JGI and other superfacility partners is that JGI data and computational challenges do not originate from standing up new capabilities, but that the continuing operation of existing services require increasingly more data and processing power over time. Many JGI services are longer lived than operational hardware and must therefore migrate data, workflows, software, and users across systems.

The Superfacility project worked with JGI on several needs, including:
- Data movement and management, which routinely transfers data between NERSC storage systems, is mediated through the JGI JAMO project, shares data with external users, or migrates workflow components to external compute resources.
- Managing a large userbase, including via Federated ID

---
[6] https://jgi.doe.gov/

● Automated analysis pipelines, including the use of Spin.

## 2.5: KSTAR

> **Key Superfacility needs:** Jupyter, Scheduling, Resiliency, API, Spin, SDN, Data movement.

The Korea Superconducting Tokamak Advanced Research (KSTAR)[7] tokamak is one of many worldwide experiments in pursuit of understanding plasma physics and engineering constraints for an eventual fusion power plant. Experiments like KSTAR often have plasma discharges every 15-20 minutes, leaving a short window of time to analyze the results and plan the next discharge. With limited on-site compute resources, this can be especially challenging for quickly understanding large datasets and computationally expensive analyses. HPC centers like NERSC can provide important computational resources to enable inter-shot data analysis, simulation, as well as AI-based decision making to help scientists make more efficient use of valuable experimental runtime at fusion research facilities like KSTAR and eventually ITER.

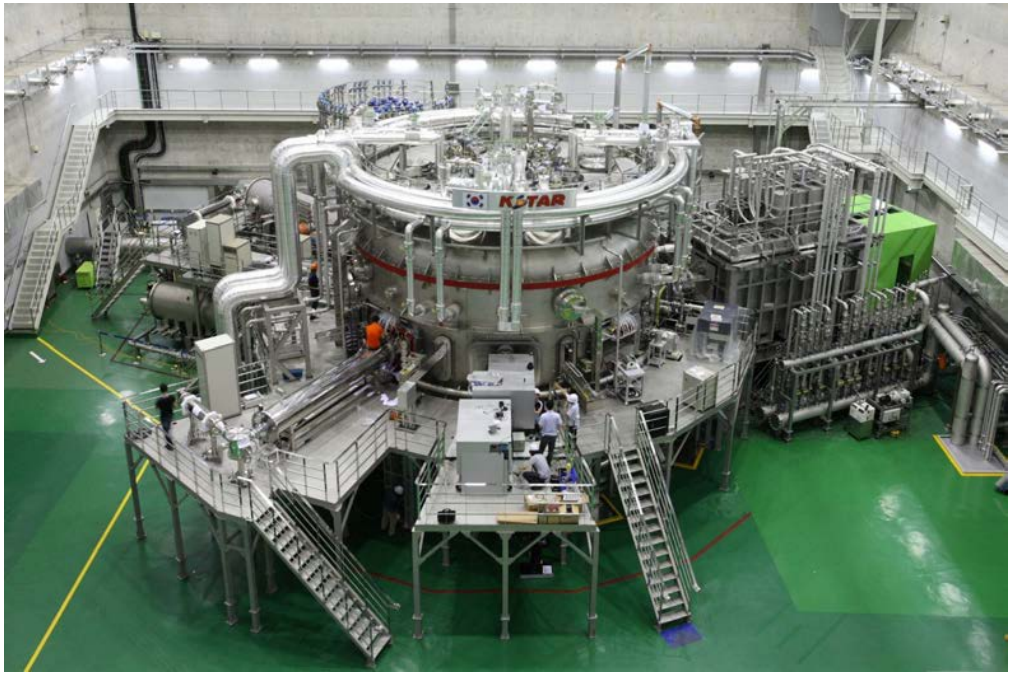

*Figure 6. The KSTAR tokamak in South Korea.*

The team is developing the Delta[8] framework which performs near-real-time data streaming via ADIOS2[9] from the KSTAR experiment in Korea to NERSC. It receives and converts the data without ever storing to disk and performs parallelized GPU-based analysis of electron cyclotron emission diagnostic data (just one of many potential datasets). Once the analysis is complete,

---

[7] https://home.kepco.co.kr/kepco/EN/G/htmlView/ENGFHP006.do?menuCd=EN070706
[8] https://github.com/rkube/delta
[9] https://adios2.readthedocs.io/en/latest/index.html

the data can be viewed on an interactive, web-based dashboard hosted by NERSC's Spin service.

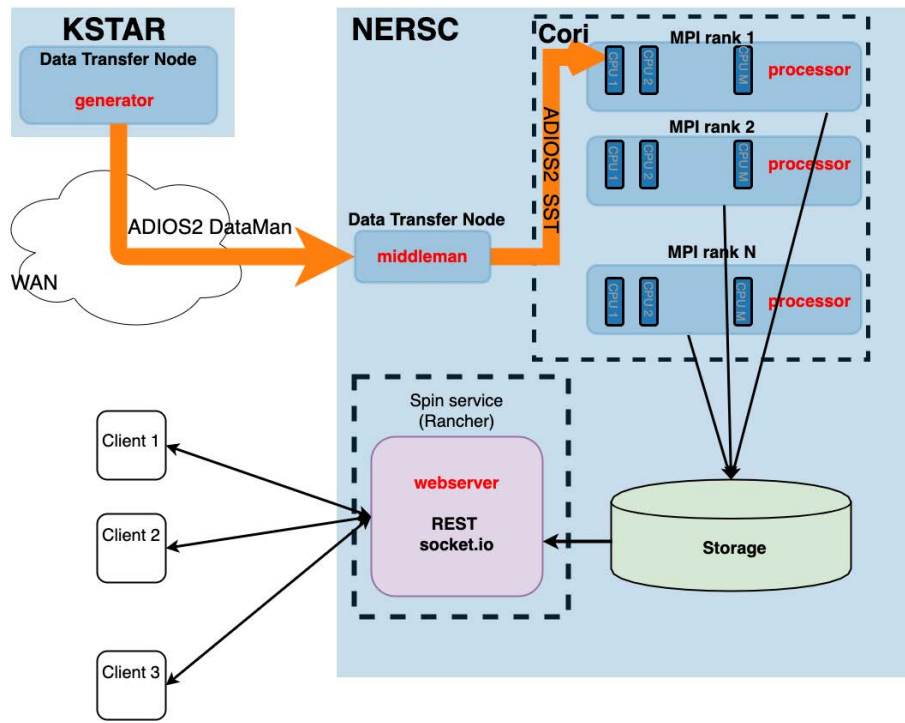

*Figure 7. KSTAR data movement and analysis pipelines (Image from https://github.com/rkube/delta)*

KSTAR joined the Superfacility project as a key science engagement in 2020, as it became clear that their framework was uniquely challenging to the NERSC infrastructure. Their requirements included:
- Support for streaming data from KSTAR directly to a compute node, via SDN
- Coordination of automated data movement and analysis, including use of Spin
- Near-real-time data analysis, including interactive analysis using Jupyter.

## 2.6: LCLS

**Key Superfacility needs:** NESAP, Policies, Jupyter, Scheduling, Resiliency, Federated ID, API, Spin, SDN, Self-managed Systems, SENSE, Data movement, Data management, HDF5.

The Linac Coherent Light Source (LCLS)[10] is located at SLAC National Accelerator Laboratory. It produces very intense femtosecond pulses of x-ray light that can be used to produce molecular movies – capturing the movement and evolution of microscopic systems in real time. LCLS has several end stations that conduct different experiments. Some of those experiments (operating roughly 5% of the year) will produce more data than can be handled locally. The data acquisition throughputs at the LCLS are expected to grow between two and four orders of magnitude by 2026 as the upcoming LCLS upgrades, LCLS-II and LCLS-II-HE, will increase the beam pulse rate from today's 120 Hz to 1 MHz. This is further complicated by the nature of XFEL experiments; substantial data analysis is required in near-real-time (usually within a few minutes after a run is completed) to effectively steer the experiment and make optimal use of limited beam time. For these high data-rate experiments, LCLS has turned to NERSC to gain access to the necessary computational resources required by live data processing.

---
[10] https://lcls.slac.stanford.edu/

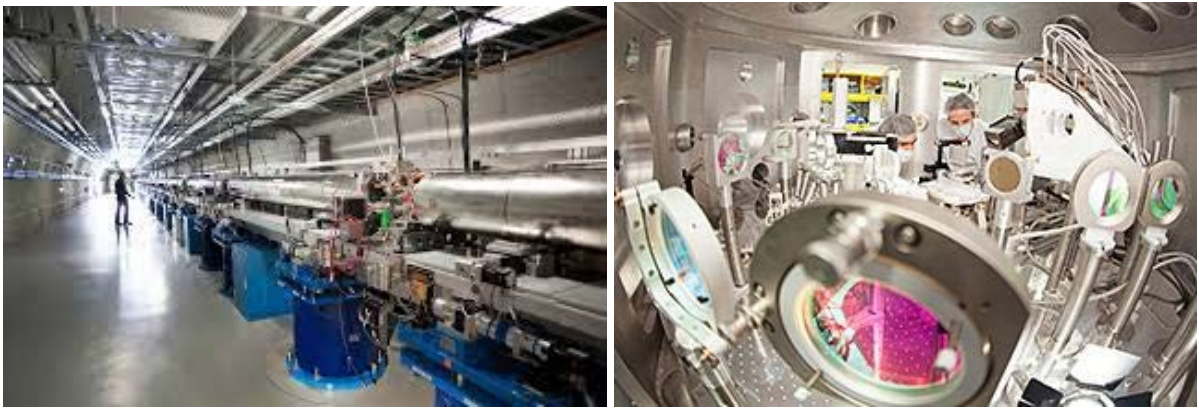

*Figure 8. The LCLS accelerator (L) and an LCLS end station (R).*

LCLS has partnered with NERSC for many years to develop the capabilities needed to support near-real-time large-scale data analysis. In partnership with the Superfacility project, we developed solutions to several of their requirements, including:
- Analysis code that can scale to exascale systems, via NESAP and the Exascale Computing Project (ECP)
- Near-real-time computing support for data analysis
- Automated data movement and analysis, using the API, Spin, SENSE and SDN
- Optimized IO in HDF5.

## 2.7: LZ

**Key Superfacility needs:** NESAP, Policies, Jupyter, Scheduling, Resiliency, Federated ID, API, Spin, Data movement, Data management.

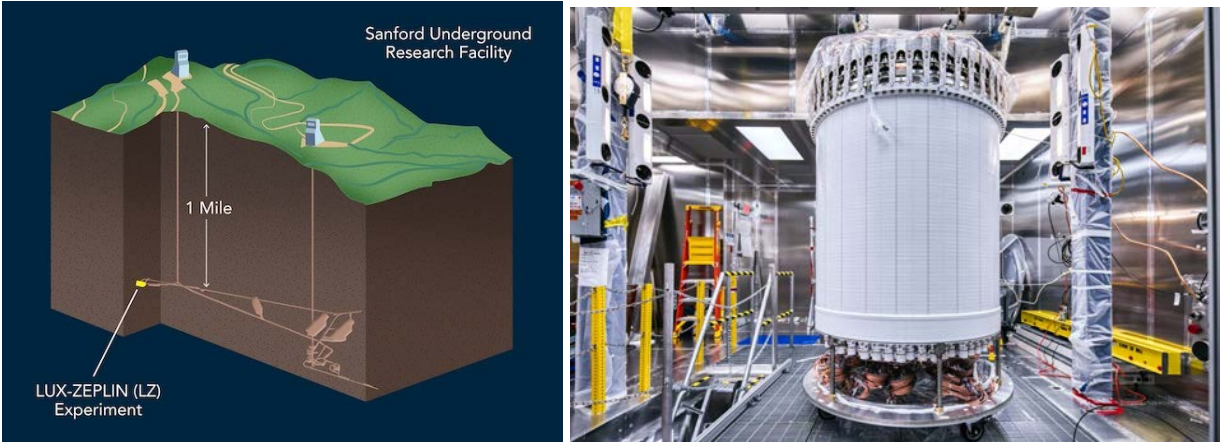

*Figure 9. Location of the LZ detector (L) and the detector itself (R).*

[LUX-ZEPLIN](https://lz.lbl.gov/)[11] (LZ) is a dark matter direct-detection experiment looking for signals from dark matter particles colliding with atoms of supercooled liquid xenon. To catch these exceedingly rare events (none have been observed so far, and no one knows if any will ever be seen), the detector sits a mile underground at the Sanford Underground Research Facility (SURF) in South Dakota, protected from cosmic rays.

LZ began operation in early 2021, using NERSC as the primary data center for data processing, calibration, and testing. LZ operates 24/7, and prompt data analysis is required to monitor the detector health and the data quality. This places requirements on near-real-time computing, workflow orchestration, and resilience, including:
- Porting simulation code to GPUs via the NESAP program
- 24/7 near-real-time automated data movement and analysis while the experiment is taking data, coordinated via Spin
- Resiliency when NERSC is unavailable.

## 2.8: NCEM

> **Key Superfacility needs:** Policies, Jupyter, Scheduling, Resiliency, Federated ID, API, Spin, SDN, Data movement.

The National Center for Electron Microscopy (NCEM)[12] facility within the Molecular Foundry at Berkeley Lab recently installed its 4D camera, which outputs data at 480 Gbit/s, resulting in single data sets of 700 GB acquired in about 15 seconds. These are orders of magnitude larger than current data set sizes at the center, and analysis/storage of these data were difficult to impossible using local resources. The superfacility capabilities implemented by NERSC provided a way for this user center to utilize HPC resources for a data-reduction pipeline for this camera.

---

[11] https://lz.lbl.gov/
[12] https://foundry.lbl.gov/about/facilities/the-national-center-for-electron-microscopy-ncem/

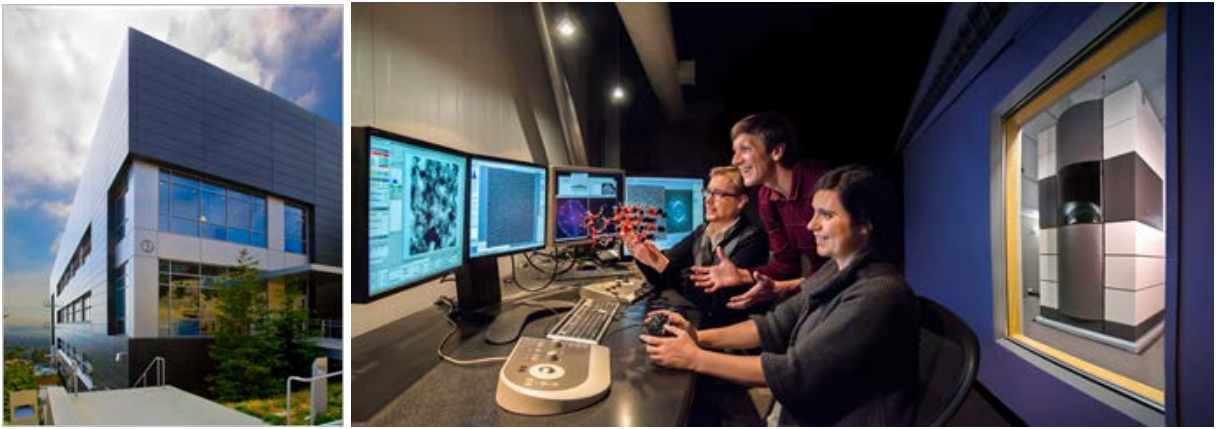
*Figure 10. The NCEM facility.*

NCEM's requirements are based around enabling near-real-time analysis of large datasets:
- In early stages, NCEM was streaming datasets directly to compute node memory, using software-defined networking (SDN) and an extension of the NERSC network directly to the NCEM instrument. This was a valuable experiment, but ultimately an unsustainable option from the security perspective.
- Now, using SDN, NCEM is transferring datasets directly to the Cori burst buffer (SSD storage layer) for analysis by compute nodes
- Automation of data movement and management via the API
- Subsequent analysis of datasets via Jupyter notebooks with specialized HPC backends.

# 3: Technology

Based on our survey of requirements across the project engagements described above, we identified several generalized capability areas in which to focus our technical work. In some cases, this work was already under way, either as an independent innovation for the center or in support of a particular project; however, in all cases the work is better informed if driven by a diversity of research use cases. Our primary aim in the project was to broaden and sufficiently generalize requirements so that the technology we developed would be reusable across current and future science cases.

The areas of technical work we identified, grouped into four classifications, are listed here:

- **User-facing tools and policies**
  - *Scalable code*, via the NESAP program, to help user codes run effectively at scale on Perlmutter

- ○ *Outreach and documentation*, to ensure all NERSC users learn about our work and how to use the tools we have developed in the project
- ○ *Policies,* to better accommodate the kinds of workflows and user communities represented by these science teams
- ○ *Jupyter*, for productive and scalable notebooks for data analysis
● **Scheduling and Middleware**
  - ○ *Advanced scheduling*, to transcend the traditional batch computing paradigm and enabling preemption, real-time jobs and similar time-sensitive computing
  - ○ *API into NERSC*, to allow easier automation and integration of HPC resources with workflow orchestration systems and science gateways
  - ○ *Federated Identity*, for better linkage between identities and accounts at disparate instrument and HPC facilities
  - ○ *Spin*, to support user-managed science gateways, databases, and other network services on-premises with access to HPC resources
  - ○ *Workflow resiliency*, to better accommodate on-demand work and workflows that span multiple instrument and/or HPC facilities.
● **Automation and Networking**
  - ○ *Software-defined networking (SDN),* for API-based local and High Performance Network provisioning and control
  - ○ *Self-managed systems*, for a future perspective on autonomy, self-healing, and an AI-driven approach to HPC systems management
  - ○ *SENSE*, for API-based WAN network provisioning and control.
● **Data Management**
  - ○ *External and internal data movement*, for easier and more performant data movement into NERSC and migration between storage tiers (flash, disk, tape)
  - ○ *Data dashboard*, to increase visibility and high-level management of data
  - ○ *HDF5*, for intrinsically richer function and description (metadata) in data sets.

The resulting project framework means many of the engaged projects have requirements in most of the technical areas. This organization allowed each technical area to be informed by real project requirements and timelines across a number of science disciplines, and for all to be loosely coupled under the umbrella of the Superfacility project goals, but also for each to work relatively independently. Developments in each of the technical areas are described in the sections below.

## 3.1: Scalable code development (NESAP)

**Science teams supported**: ALS, DESC, DESI, JGI, LCLS, LZ

The NERSC Exascale Science Applications Program (NESAP) is a collaborative effort in which NERSC partners with various stakeholders to prepare for future computational architectures and address challenges that arise from porting existing codes to these architectures.

One of these challenges is how to enable existing codes to make good use of GPUs. Even though GPU readiness is a central objective of NESAP, the project also works on other aspects of superfacility workflows, such as containers, workflows, and data movement (into and within NERSC). NESAP allows for a much closer collaboration between NERSC and users, giving code development teams access to NERSC staff time, new hardware (as it is being commissioned at NERSC), and NESAP postdocs. NESAP codes cover a wide range of science domains, from simulation to data analysis, and machine learning. Some NESAP teams are supported by ECP, including ExaFEL, an analysis framework for LCLS. The ALS, DESI, JGI, and LZ projects (see Section 2) also all have NESAP projects in addition to the Superfacility collaboration. The table below shows a snapshot of the performance improvements achieved by some of the NESAP teams.

| Name | GPU Acceleration<br>System comparison | SSI |
|---|---|---|
| TomoPy<br>ALS | 97x (time to process 24 slices)<br>Edison vs Perlmutter Phase 1 (A100) | 26.59 |
| ExaFEL<br>LCLS | 2256x (time to simulate 1 image)<br>Edison vs Perlmutter Phase 1 (A100) | 214.76 |
| DESI | 25x (frames per node-hour)<br>Edison vs Perlmutter Phase 1 (A100) | 6.98 |
| ExaBiome<br>JGI | 13.5x (PASTIS benchmark)<br>Cori Haswell vs Perlmutter Phase 1 (A100) | 9.61 |
| FICUS<br>JGI | N/A | 5.37 |
| LZ | N/A | 36.46 |

*Selected performance highlights for NESAP for Data projects (of a total of 52 NESAP projects). All of these projects aim to accelerate experiment data analysis using NERSC's systems. Since raw speedup numbers do not reflect the increased capabilities of a new hardware (e.g., more available cores), we also show the scalable system improvement (SSI)[13] – a metric that allows comparisons between systems.*

NESAP gives teams access to NERSC and vendor expertise in the form of regular hackathons and NERSC staff time. Hackathons are held regularly four times a year. They allow teams to work intensively on code acceleration problems together with mentors from the ASCR HPC facilities and from vendors, and they are often the starting point of new features or code redesigns. Furthermore, NESAP teams are assigned a NERSC liaison, plus NESAP postdocs if a suitable candidate is matched with the NESAP project. In addition to staff expertise, NESAP teams were given early access to Perlmutter and priority queues on Perlmutter and Cori GPU. This has allowed these teams to be some of the first users to test their codes on GPUs (first NVIDIA V100s, and later A100s) and Perlmutter.

---

[13] https://www.nersc.gov/research-and-development/benchmarking-and-workload-characterization/ssi/

This close level of engagement with users benefits the domain scientists as well as NERSC. By providing expertise and staff time, domain science codes are able to better leverage new hardware. This is evident not only from the speedups listed in the table above, but also in enabling new capabilities, including, but not limited to, these projects:
- ExaFEL/LCLS has ported device code to Kokkos, which has enabled ExaFEL codes to be portable across architectures
- DESI has included containers and improved their data analysis pipeline (and optimized job sizing). This has eliminated bottlenecks in their data processing pipeline.

Granting early access to NESAP teams has allowed NERSC staff to better stress-test the Perlmutter system under realistic workloads. These engagements are therefore useful at catching bugs – for example:
- DESI and ExaFEL/LCLS revealed a bug in Cray-MPICH when applications call fork()
- LZ and DESC have been early testers for CVMFS on Perlmutter
- ExaBiome/JGI has been an early user of GASNet on Perlmutter.

## Gaps/Next Steps

As the NERSC workload evolves, the NESAP program also needs to evolve. The NERSC-10 system will be designed specifically to support complex workflows, and NESAP will need to adapt to the increased complexity this presents to users. While future NESAP efforts will still include kernel performance tuning, we expect that the computational performance will no longer be the biggest factor determining the "time to science."

One of the biggest questions we are considering is how to quantify total workflow performance. Workflows contain many steps – data movement, perhaps multiple stages of compute (on heterogeneous hardware), coupled simulation and data processing, IO-intensive stages, and combining data from multiple sources. This expands the scope of NESAP, in addition to kernel performance optimization, to account for a more holistic view of how the different workflow components interact. It is hard to profile a fully end-to-end workflow and identify bottlenecks when these workflow stages may run at very different timescales, and a total workflow campaign may last months. It is even harder to define a single metric that encapsulates the total time to scientific insight by which we can measure success. A future NESAP program will need to carefully define success in a way that takes into account these factors – probably with multiple success metrics that map to different workflow stages and the specific hardware of the NERSC-10 system. Based on our experience with the superfacility workflows, we anticipate that future NESAP staff will include a broader range of experts at NERSC who are qualified in the different workflow components.

Recruitment and retention of post-docs for the NESAP program has always been a challenge and will continue to be so in the future. We have been able to hire very strong post-docs, and many do stay on to work as NERSC staff long-term, but this is a very competitive sector and others do leave to pursue other opportunities. This will need to be considered in the future

NESAP program; a broader range of experts to support NESAP workflows will help alleviate hiring gaps.

## 3.2: Outreach and documentation

> **Science teams supported**: ALS, DESC, DESI, JGI, KSTAR, LCLS, LZ, NCEM

In order to sustainably support the technologies developed as part of the Superfacility project, we developed documentation and examples for using these new tools and services. One example of documentation developed in the project is written instructions for, including examples of, using the API. This information is hosted on the user-facing NERSC [documentation pages](#).[14] Interacting with HPC systems has not traditionally been done via APIs, so it is imperative that we communicate to users how and when to use the SF API in order to facilitate using the tool for automated workflows. In the process of documenting example use of the API, we wrote the commands into a usable, shareable Jupyter notebook. We presented demonstrations of the API to several groups (e.g., the ALS computing team) and distributed the Jupyter notebook to interested users following the demonstrations. This combination of static plus interactive documentation and examples is a model we are increasingly moving toward with our documentation at NERSC.

Beyond the Superfacilty API, we developed a list of best practices for NERSC users conducting experimental and observational science work, hosted on the NERSC [documentation web pages](#).[15] These curated instructions build on a foundation of standard best practices for all NERSC users and incorporate the specific needs of experimental and observational scientific workflows. Topics on the page include data management and sharing, file systems, and job submission.

The superfacility model is of increasing importance to the HPC community, and is reflected in the number of conference sessions and workshops dedicated to the topic. The Superfacility team has driven this community-wide discussion via conference submissions, panel discussions and feature talks at many venues including SC, ISC, PASC, CUG, SciPy, PEARC, and multiple DOE workshops. Outreach highlights include two "State of the Practice" talks [1,2] at SC20 as well as one paper presented as part of the SC20 XLOOP workshop [3]. We also presented several demonstrations at DOE (and other) booths at SC19, SC20, and SC21.

We have also focused on outreach to our user community, to inform them of the new tools and capabilities we have developed, and help them get up and running with them. One particularly successful example of this was the [Superfacility Demo Series](#)[16], a series of webinars hosted by NERSC in 2020, featuring talks and demonstrations from NERSC and science partners' staff.

---

[14] https://docs.nersc.gov/services/sfapi/
[15] https://docs.nersc.gov/science-partners/bestpractices-eod/
[16] https://www.nersc.gov/research-and-development/superfacility/#toc-anchor-3

These highlighted the ways in which NERSC staff have partners with our Superfacility science engagements to develop tools that are having real impact on their work.

**Gaps/Next Steps**

Outreach to different research groups has varying degrees of success. Within large scientific collaborations like the ALS, for example, there is a wide range of expertise with a differing amount of focus relating to the use of supercomputing resources like NERSC. It became apparent through the course of outreach efforts that a science team cannot be approached as a monolith. This was identified as a gap in our approach to directly engaging with science teams, and resulted in a more focused approach. In the near future we will connect with leaders or representatives of groups within science teams to better understand the groups' technical readiness for using NERSC and associated superfacility technologies. This will allow us to reach the scientists who will actively interact with NERSC and focus our training efforts appropriately.

As our user base continues to expand as we support more large collaborations from experimental facilities, we will need to scale our user support accordingly to empower users to help themselves. There are many different learning styles; not everyone learns best by reading documentation, although that is the dominant mode of user communication and support. As we reach more scientists via the Superfacility model, we will need to expand our training with more videos and hands-on worked examples, eg via Jupyter notebooks. We will also focus more on Superfacility topics in the NERSC trainings and workshops.

## 3.3: Policies

**Science teams supported**: ALS, DESC, DESI, JGI, KSTAR, LCLS, LZ, NCEM

The Superfacility project has brought interesting new considerations of policy design and communication to NERSC. Policy is a tool to set NERSC and DOE ASCR priorities, set user expectations and effectively communicate and manage resources.

The Superfacility project included many areas of technical work that required policy changes to support them. For example, through our advocacy for greater resilience to outages to support superfacility science teams, we were able to motivate changes in the procedures around facility and system maintenances that resulted in systems and auxiliary services staying online (see Section 3.7 for details). Another example of policy changes driven by the Superfacility project were around queue policies and system scheduling for real-time workflows. We devised policies that could support real-time workflows, while also keeping utilization high (see Section 3.7 for details).

The Superfacility project also expanded the purpose of policy to the interface between NERSC and other institutions' governance and decision making. Each potential partner institution has its own unique set of requirements, goals, and policies; successful Superfacility collaboration requires compatible policies across each participating organization. The policy system works best when those policies are easy to find and well written so questions can be answered quickly.

A concrete example of these concepts in action is the intersection of the Superfacility federated identity (FedID) component and security policies. The system implementing Superfacility FedID must balance DOE security requirements, NERSC security policy, the technical capabilities of available authentication technologies, and the individual security policies of other potential partners. There will inevitably be misalignments of risk tolerance between institutions but as we increase the visibility and quality of our offerings it becomes more likely that we can convince other organizations to modify their policies and grow our pool of partnerships.

## Gaps/Next Steps

While the Superfacility project has identified and advocated for a number of policy changes at NERSC, a number of open policy issues remain. We detail key areas where policy needs to be reevaluated below.

The Superfacility project science engagements tend to be large collaborations, often with several hundred NERSC users – they are some of the largest projects at NERSC. These users often have different roles within their organizations; some still need to SSH directly to NERSC systems, but many could perform their role with limited access such as only viewing or transferring data, connecting to Jupyter, or maintaining services hosted in Spin. Today, NERSC only has one policy definition of user, so each of these use cases must be served with full user onboarding process, security vetting, and results in full access. It would be useful to rethink how we define users and projects at NERSC to better reflect the organization of these large collaborations. For example, introducing user roles with diminished access (e.g., only data access or run pre-defined analysis routines) could allow us to lift security or policy constraints that would make such users much more convenient for the user themself, the collaboration leadership, and NERSC support staff.

Other outstanding policy issues require NERSC to rethink how we allow and enable access to our systems in the long term. Currently, time at NERSC (and other computing facilities) is allocated on an annual basis with no formal guarantee of renewal. (Many projects informally discuss the need for longer term allocations with program managers; however, with program manager turnover this can become a challenge for projects.) Long-running science experiments need to be able to rely on access to NERSC for the duration of their project, otherwise they risk not being able to process their data in the future. Guaranteed access to NERSC resources over a period of years would also justify the large amount of valuable scientist time that is sunk into setting up and optimizing compute infrastructure for NERSC. In an ideal world, this workflow infrastructure would be able to operate equally well at other HPC centers, but that is not yet the case (see [Section 3.7](#)).

Supporting experiment teams who require real-time access to NERSC systems opens the issue of how to prioritize our workload. At present, the real-time computing needs are relatively small (see [Section 3.7](#)), but as they increase we will need a process to decide who gets priority access to resources and how to choose a balance between urgent projects and the disruption they may cause to the traditional batch HPC workload.

This also has implications for how we define our success metrics. NERSC currently reports both utilization (tied to predictable, efficient resource scheduling) and "capability" job metric (based on the percentage of hours at NERSC where jobs use a large fraction of the machine). The Superfacility model suggests we should expand the notion of what *capability* means at an HPC facility to better support a wider range of workloads. Going forward we need to thoroughly understand the trade offs between utilization and real-time support and devise policies which resolve conflict between the two. A thorough assessment of HPC metrics is essential as we expand into new models of supporting science.

In summary, top areas for new policy consideration or reevaluation to support Superfacility like workloads include:
- Multi-year allocations
- Expanding user definitions and policies to enable stratification of access and convenience; many users don't need SSH, they may only need data transfer or a limited NERSC resource API
- Prioritization of real-time versus batch scheduled work
- NERSC's capability metric and other traditional HPC facility metrics
- Sharing and dissemination of policies.

## 3.4: Jupyter

**Science teams supported**: ALS, DESC, DESI, LCLS, LZ, NCEM

Experimental and observational data (EOD) facility users need supercomputing to make sense of all the data they generate. But these scientists are usually newcomers to supercomputing who find traditional interfaces (e.g., CLI) to be intimidating and unfamiliar. They need a scientific software platform upon which they can build, run, capture, annotate, re-run, and share analysis workflows. Jupyter is the de facto platform for data science and AI, so it is natural for data-intensive science users like our Superfacility partners to expect Jupyter to work on a supercomputer. That is why Jupyter is a part of the Superfacility project. In just three years the number of unique NERSC Jupyter users has doubled to 2,163 (in 2021), indicating that most active NERSC users are now Jupyter users.

The Superfacility Jupyter project involved a research and development team from the Data Science and Technology Department (DST) in the Computational Research Division (CRD) working in close collaboration with the Jupyter deployment effort at NERSC. We built on partner efforts to identify gaps and pain points through requirements interviews with science teams [4], prioritizing issues common to multiple teams. These resulted in agile software projects to extend and adapt Jupyter in HPC. Deploying JupyterHub as a collection of microservices in Spin, building the constituent containers using continuous integration, and automating deployment of JupyterLab on Cori and Perlmutter made it possible for NERSC to keep up with experimental developments from the R&D team and to provide rapid feedback from the production Jupyter deployment.

Our co-development work with DST focused on developing tools to enhance the Jupyter user experience in NERSC and HPC computing environments, and deep-dive engagements with science groups to enable experimental facility workflows at NERSC. We introduced extensions to enable better file system navigation at HPC centers with large shared file systems and complex, sprawling folder hierarchies (*jupyterlab-favorites* and *jupyterlab-recents* extensions to bookmark and highlight commonly accessed files and global filesystems). Large science teams often have complex software environments that need to be maintained across the collaboration. We introduced the concept of custom project Jupyter environments with a specialized software stack and Jupyter tools. Our *Jupyter Entrypoint* service allows users to launch these custom environments with their project stack. We have developed a *Jupyterlab Slurm* extension to allow users to interact directly with the batch queue system at NERSC through Jupyter. Other user-facing capabilities we developed or integrated into the HPC environment include a service to clone curated notebooks, an extension for managing announcements to users, enhancements to the help system, and resource usage monitoring tools.

Our collaboration also had deep-dive scientific engagements with Superfacility partners to enable Jupyter-based experimental workflows. Our approach involved working closely with domain scientists to develop Jupyter notebook-based workflows, identifying key scaling bottlenecks and gaps in the interactive user experience. We iteratively developed workflows that could leverage NERSC HPC resources directly through the Jupyter interface, while interacting with the results from these jobs. We worked with science teams, including NCEM, ALS, and LCLS data-processing pipelines, to capture common Jupyter workflow patterns that can be deployed in a repeatable manner across multiple projects. We describe a case study from NCEM here as an example [5].

**NCEM Image Analysis**

Scanning transmission electron microscopy (STEM) is used for spectroscopy, diffraction, and imaging of materials. It is possible to determine many structural properties of materials using four-dimensional (4D) STEM images. Bragg Disk detection is an important part of this process and can require computationally expensive image-processing steps. Scientists at NCEM originally developed a serial version of the Bragg Disk detection algorithm using Python and NumPy, running as a Jupyter notebook, where electron microscope image data were rendered for inspection and analysis. The image processing could be easily vectorized and was a good candidate for parallelization with a tool like Dask.

By enabling a parallel implementation for the Bragg Disk detection code through Jupyter at NERSC, we were able to significantly improve run times for large data: processing a 300GB dataset went from days to a few minutes. This workflow was scaled up to run on 40 Cori nodes (1,280 workers), with results collected and visualized in the Jupyter Notebook. This type of scaling has opened the door for computational analyses that were previously impractical. One of the benefits of the 4D-STEM technique is the ability to perform multi-scale data analysis, combining atomic-scale spatial sampling with very large fields of view. Without a scaled parallel implementation, one would need to give up either the spatial sampling or the field-of-view when analyzing larger datasets.

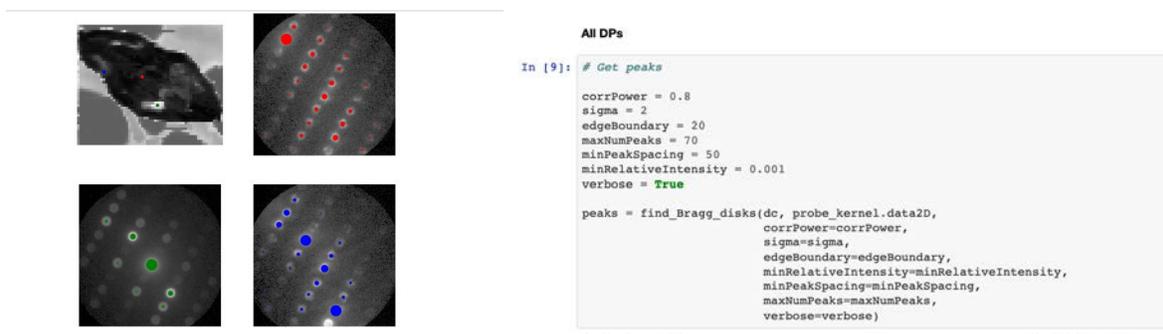

*Figure 11. NCEM Bragg Disk Detection Jupyter Workflow, which achieved significant performance gains (20-100x).*

## Gaps/Next Steps

Multiple user facilities have expressed interest in new real-time collaboration (RTC) capabilities for Jupyter notebooks that enable multiple people to interact with a notebook simultaneously (like Google Docs). To make this safe and secure enough to work at NERSC, new capabilities are needed in JupyterLab and JupyterHub, and realizing those capabilities will require very close collaboration with Jupyter developers. User facilities have also outlined a preferred paradigm for guest users that allows them to leverage Jupyter at NERSC *but without requiring their users to have actual NERSC accounts.* Role-based access controls needed for RTC may provide restricted scope that could allow this, but it also involves complex policy negotiation. This may also be tied to additional work to be done around auditing and logging user activity, and potentially enabling modes of shared access.

Given the importance of new architectures like GPUs, NERSC should investigate patterns and pain-points to effectively manage distributed Jupyter workflows on GPU systems, along with possible solutions through real-world science partnerships (and possibly a showcase demo). As containers become more tightly integrated with HPC systems, NERSC will need to continue and build on current activities to support use cases where science workflows are captured as reproducible containers.

## 3.5: Scheduling simulator

Superfacility workloads require special consideration in scheduling owing to their real-time deadline-driven nature. If an experiment is creating a significant amount of data in a bursty mannerthat needs to be analyzed quickly, fast access to resources in a timely fashion is critical. Our main goals in developing a scheduling simulator was to:

- Study the impact on queue wait times of carving out a real time partition for dedicated access to SF workloads, and its impact on other NERSC workloads.
- Use the simulator to come up with optimal approaches to support these workloads.

We analyzed existing scheduling slurm simulators, and tried to simulate a system the size of Cori to study impacts on queue performance. The overall learnings from the simulator were:

- The backfill scheduling algorithm can be simulated with great accuracy for a given workload. Simulators in many cases arrived at the similar scheduling decisions observed in real workloads.
- To obtain high accuracy, simulators experience extremely long running times. In many cases the simulator ran with similar algorithmic complexity as the actual Slurm controller after abstracting various network effects. If we included Cori's job sizes and site policies, the simulator's running times lagged behind the slurm controller by a constant factor.

Long running times meant that our original goal of using the simulator for studying the queue wait times was not viable, as simulating a production workload became computationally very intensive. For these reasons we concluded that analyzing our extensive scheduling history can yield a much better understanding of queue wait times. The Slurm simulator continues to be an interesting exercise in trying out new scheduling approaches and testing new scheduling algorithms and understanding their algorithmic complexity, but was not a viable approach to figure out how to support a workload through an existing scheduling algorithm.

## 3.6: Real-time scheduling

**Science teams supported**: ALS, DESI, KSTAR, LCLS, LZ, NCEM

One of the key (and new) features of the workloads we aim to support with the Superfacility project is near-real-time computing. Several of our science engagement teams require HPC-scale computing at NERSC to analysis their data to monitor and control their experiments as they operate (e.g., LCLS needs to see results of an experiment within minutes to decide how to adjust the experiment for the next x-ray shot), and others require short turnaround data analysis to ensure timely decision making (e.g., DESI needs to analyze nightly telescope data by morning). This is a new paradigm in scheduling work at busy HPC centers, which typically rely on a batch system to get through the queue of compute jobs waiting to run.

The solution to this problem included both policy decisions, and engineering work. NERSC identified two solutions, both of which are in use on NERSC systems today:

1. **The "real-time" QOS**

Since 2017, NERSC has supported small-scale compute needs that require very short turnaround via a "real-time" quality of service (QOS) in Slurm. Access to this QOS is tightly constrained to NERSC users who have genuine need for it (including ALS, DESI and LZ). A small set of compute nodes are held in reserve to serve this QOS. If these nodes are all occupied, then incoming near-real-time jobs are assigned a high priority in the Slurm scheduler which will allocate them to the next available compute nodes. In a system as busy as NERSC's Cori (which typically has over 2000 jobs running simultaneously) the wait for resources will be minimal (typically a few minutes).

2. **Reservations + preemption**

The near-real-time needs of experiment facilities like LCLS are projected to rise rapidly in the next five years, when LCLS reaches full capacity they will regularly require hundreds of petaflops of compute to monitor and guide their experiments. This scale is very hard to supply via the real-time QOS, so in collaboration with SchedMD we have developed a system of reservations to provide the required node availability, and still maintain high utilization of our systems. LCLS needs this level of compute capacity for specific experiments which happen at scheduled times in the year; these are not random, unpredictable needs. With advanced warning, we can reserve compute nodes for the duration of the LCLS experiment so they can surge to NERSC whenever they need to.

The challenge in this scenario is that experiments don't always run smoothly. During a shift at an experiment, equipment may fail, or samples may be hard to adjust - so for much of the time the reserved compute nodes may be sitting idle. To avoid this and maintain good use of our compute resources for science, we have developed the capability (with SchedMD) to enable preemptible jobs to run in reservations. This means that any flagged compute job can run on nodes set aside for a reservation, with the expectation that the jobs will be canceled within a given time frame (generally a couple of minutes) so that the nodes can be given to the experiment team when they have compute jobs flowing to NERSC (see Figure 12). This capability was funded by NERSC via NRE with SchedMD, and was deployed in 2021.

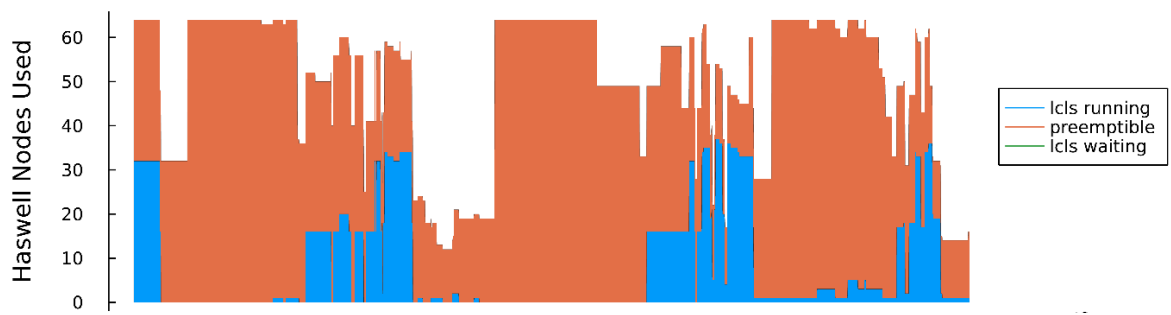

*Figure 12. Example of compute usage of a reservation for a running experiment, and how near-real-time data analysis jobs (blue) can coexist with preemptible jobs (orange). 64 nodes (2048 cores) were reserved, to which near-real-time jobs have priority access.*

**Gaps/Next Steps**

In the next five years, we expect an increasing amount of our workload to come from short-notice, rapid-turnaround compute jobs from experiment facilities. Our system of reservations plus preemptible jobs is working well at relatively small scales - it is unclear whether this will scale to jobs that require a significant fraction of the machine. Work will be needed to appropriately incentivise a preemptible workload to fill the gaps in reservations. This will include increased use of (and support for) checkpointable applications.

An outstanding problem is how to handle incoming urgent compute demands that do not have a reservation (for example, if a supernova goes off nearby, several HEP experiments will need quick turnaround data analysis, and supernovas are entirely unpredictable). A large pool of general workload that is preemptible (e.g., via user or perhaps system checkpointing) may be an option, but significant work is still needed to identify a technical solution.

 A key question for rapid-turnaround workloads is, who gets priority? How many demands for near-real-time computing can we handle at any one time? How do we ensure non-urgent compute jobs also have a reasonable turnaround? When do we say no to an incoming request? These policy questions will be closely debated in the next years as we see more requests for real-time computing resources.

## 3.7: Workflow resiliency

**Science teams supported**: ALS, DESC, DESI, JGI, KSTAR, LCLS, LZ, NCEM

Several of our science engagement teams have time-sensitive computing needs. For example, LZ needs to monitor the health of their detector and data quality 24/7, DESI needs to process data coming from the telescope every night, and ALS needs access to data stored at NERSC during experiment shifts that are scheduled far in advance. The Superfacility project addressed workflow resilience in two ways:
1. Making NERSC more resilient to outages and maintenances
2. Helping users develop a more resilient workflow system.

In June 2020, NERSC held a resilience summit to discuss how to better support resilient workflows, which has resulted in plans for how to improve facility, system and user workflow resilience. The Superfacility science teams provided the driving use cases and continues to motivate work to improve NERSC's resilience.

**Making NERSC more resilient**
The following table lists the infrastructure required by each science team to remain productive during power work or maintenances. Based on these needs, NERSC is now able to keep many services up and running during outages (in green), using generator power to support the highest priority services. We are also exploring how to keep a small set of compute nodes available on backup power. Some of examples of the ways the facility has improved resilience include:
- Added mechanical equipment capacity to at least N+1 redundancy or better:
    - 3 new cooling towers - 7 total
    - 3 new pumps on each of two piping loops - 5 total on each loop
    - 3 new heat exchangers - 4 total (eliminated the primary single point of failure)
    - 3 new air handling units (AHUs) - 7 total
- New equipment includes improved protection from environmental hazards - moisture, smoke, and dust
- Added an additional AHU onto standby power system, providing 2 units to improve air flow capabilities during power outages and wildfire smoke conditions
- Augmented and improved our weather and particulate sensor array to provide informed decisions during inclement environmental conditions
- Added physical and cyber security improvements to reduce exposure to unexpected system changes or unauthorized access.

|      | Network | DTN | Spin | /global/ common | CFS | HPSS | Login node | Compute nodes | Jupyter |
|------|---------|-----|------|-----------------|-----|------|------------|---------------|---------|
| ALS  | ✓✓✓     |     | ✓✓✓  |                 | ✓✓✓ | ✓✓   | ✓✓         | ✓             | ✓       |
| DESC | ✓✓✓     | ✓✓✓ | ✓✓   | ✓✓✓             | ✓✓✓ |      | ✓✓         | ✓✓            | ✓       |
| DESI | ✓✓✓     | ✓✓✓ | ✓✓✓  | ✓✓✓             | ✓✓✓ | ✓✓   | ✓          | ✓             | ✓       |
| JGI  | ✓✓✓     | ✓✓✓ | ✓✓✓  | ✓✓✓             | ✓✓  | ✓✓   |            | ✓✓            |         |
| LZ   | ✓✓✓     | ✓✓✓ | ✓✓✓  | ✓✓              | ✓✓✓ | ✓✓   | ✓          | ✓             |         |

**Helping users develop more resilient workflows**
We developed and demonstrated capabilities to transfer NERSC-based workflows to other sites. For simple pipelines, we showed how Jupyter and conda apps can be ported to other sitesï[17]. For more complex pipelines, we initiated two projects: an ALCC-supported project to demonstrate complex workflows at the ASCR facilities, and an LDRD-funded effort to port DESI and LZ pipelines from NERSC to the LBNL Lab IT cluster. The lessons learned from this work are being documented[18][6] and as a result we have developed a Resilience Policy[19] to advise science teams on how to develop more resilient workflows.

---
[17] https://github.com/NERSC/c2d

## Gaps/Next Steps

While achieving 100% availability for all of NERSC's systems and services is unlikely due to necessary system and facility maintenances, we strive to architect our systems to minimally impact users, with the goal of architecting systems and services that appear to be always available, from the user's perspective. Rolling upgrades to compute node software are one aspect of this; users see that some nodes are momentarily unavailable in the compute pool, but no workload is actively disrupted while the updates are going on. NERSC is already able to do this for some storage systems and some Cori compute node updates, and Perlmutter will also have this capability. Hiding hardware failures is harder to do – e.g., if a compute node goes down unexpectedly the application running on it will fail – but the impact can be mitigated by (for example) automatically re-queueing the compute job, or supporting system checkpoint/restart. These options are actively being explored at NERSC.

 A harder problem is how to protect user workflows from disruptions in the system, for example network contention or storage systems performance slowdown (e.g., because of a user's job hogging resources). This may not cause an application to crash, but can severely impact its performance. Methods to mitigate this via hardware and system software is a topic under exploration for the NERSC-10 system.

For users to be able to build resilient workflows, it will be important to expose performance information about current conditions in the various systems at NERSC (storage, compute, network), and within ESnet. This information should be exposed in a such a way that it can be acted upon by an automated workflow – e.g., via an API call that allows a user to see the current I/O load on a particular storage system, and to launch or delay their data processing accordingly. This is also a topic being explored in the NERSC-10 project. Giving users the data about system status empowers them to make their own decisions about how to use an HPC center, and gives them valuable insight into system performance.

Some failure modes are impossible to prevent; for example, a large-area power outage or natural disaster. In these cases, a time-sensitive user workflow needs to be able to switch to a different computing site at short notice. (Workflows may also want to do this without the prompt of a disaster, to take advantage of specialized architecture or data at different sites). Currently this is very hard - almost impossible. A significant investment in infrastructure, tools and support is needed to truly enable cross-facility workflows. This problem is currently being considered by the DOE SC Integrated Research Infrastructure effort, and other grass-roots projects are attempting to link computing centers across agency, university and commercial boundaries. We expect that this will be a major area of work and investment for scientific workflows in the next 5-10 years.

---

[18] https://crossfacilityworkflows.github.io/BestPractices/
[19] https://www.nersc.gov/users/policies/resiliency-planning/

## 3.8: Federated identity (FedID)

**Science teams supported**: ALS, DESC, JGI, LCLS, LZ

Every discussion of distributed workflows eventually touches on the issue of authentication and the complexity and inconvenience brought on by having multiple passwords and accounts across instrument and compute facilities. The work in this area was intended as a first step at addressing these problems by a) embracing modern approaches to identity management, b) leveraging modern distributed authentication protocols – OpenID Connect (OIDC) and the Security Assertion Markup Language (SAML) – and c) applying them to science use cases.

*Federated identity* (FedID) is the concept of establishing a trustworthy association between an individual's digital identities; *federated authentication* is the mechanism that uses these associations to authenticate to systems and services. As of the writing of this report, NERSC is piloting federated identity with Lawrence Berkeley National Lab, allowing users with LBNL accounts to link them to their NERSC account, then use them to log in to core NERSC systems. During the pilot, the systems at NERSC that support federated authentication are Iris (account and allocation management), ServiceNow (support tickets), the NERSC web site, and Spin. In the coming months, the FedID capability will be expanded to hundreds more institutions and additional services at NERSC.

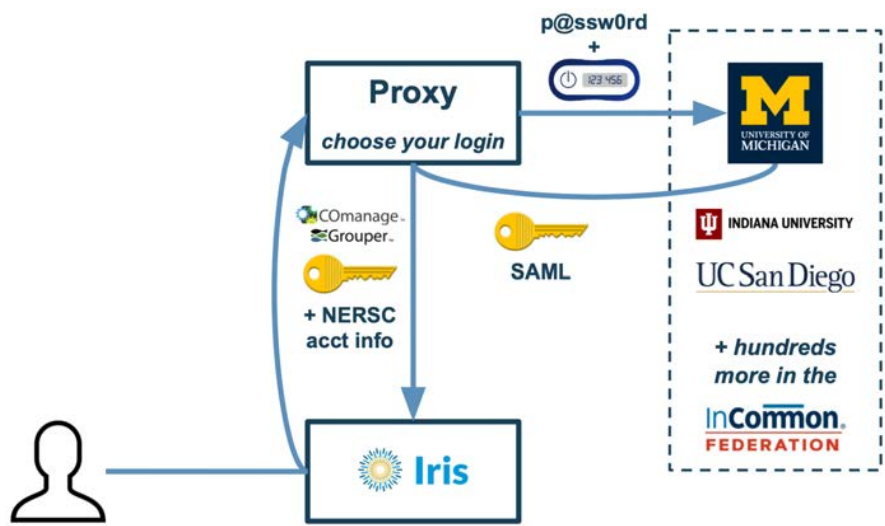

*Figure 13. Federated authentication login flow.*

The distributed nature of federated identity requires that trust be placed in additional technology and other organizations and their security practices. NERSC underwent a full study comparing its conventional, self-contained authentication versus the federated approach. Ultimately, it was decided that security requirements would be satisfied by:

- Scoping its implementation to DOE-managed facilities and members of the InCommon Federation, an established trust network of higher education and research institutions created to enable federated identity in the academic community
- Restricting any non-DOE participating institutions to those that comply with SIRTFI (a set of best practices for organizational security) and the REFEDS MFA Profile (a protocol extension that signals when users perform multi-factor authentication.

**Gaps/Next Steps**

The initial pilot of federated identity for the subset of NERSC users that are affiliated with LBL has shown that we can provide an enrollment process that is easy to navigate, that the impact of additional support tickets is minimal, and that if made available, many users will take advantage of it (over 300 LBL users have signed up in the first 6 weeks of availability, including users from ALS, DESI, DESC, JGI, LZ and NCEM). There is still work remaining to make this more broadly useful:
- Expand eligible users to members of other DOE labs, and eventually to all members of the InCommon identity providers that meet our security requirements (as described in the project plan)
- Expand the applications that can utilize our federating identity framework, by enhancing our login proxy to work with the OpenID protocol, in addition to the SAML protocol.
- Develop a web application that can provide SSH certificates via a federated login
- Improve the user experience during the multi-factor authentication (MFA) step-up process. Currently, if the user's login provider doesn't signal to NERSC that a multi-factor authentication was used, we "step-up" the authentication by issuing an MFA challenge for a NERSC token. The current software doesn't limit the number of MFA challenges a user will receive. Adding the concept of an MFA session to the proxy would improve the user experience by reducing the number of MFA challenges to the one per day they have grown accustomed to in our legacy login process
- Leverage capabilities made available through the Office of Science Distributed Computing and Data Ecosystem project. Developments out of this project could streamline the process of managing the members of science projects that work across DOE facilities, as well as streamline the process of enrolling a federated identity at NERSC.

## 3.9: API access into NERSC

**Science teams supported**: ALS, DESC, JGI, LCLS, LZ, NCEM

As automation is increasingly required for complex workflows, and science gateways become an increasingly popular digital interface for user communities, it is important for HPC centers to provide a modern and secure means for software to interface with batch workload managers, storage systems, and other supercomputing facility resources.

The NERSC API is a modular application programming interface that allows science teams and experimental facilities to authenticate to NERSC systems and perform their work. The API was developed with input from several science teams currently using NERSC resources via mostly home-grown software (particularly ALS, DESC, LCLS and NCEM). Through the interview process we identified common abstractions which then became API components.

Each API component is a REST web endpoint which can be accessed from any programming environment. The main API endpoint allows users to test API calls right from the browser and also serves the Swagger-style API documentation.

For user authentication the NERSC API uses OAuth2, an industry standard familiar to most programmers. OAuth2 is an authentication protocol only and does not tie users to any programming language or environment. Once authenticated, NERSC users may then call the API. The following API components are available to our users:

- **account**: Get accounting data about the user's projects, roles, groups and compute and storage allocations
- **compute**: Run batch jobs and query job and queue statuses on NERSC compute resources
- **status**: Query the status of NERSC component systems
- **storage**: Transfer files between Globus endpoints
- **utilities**: Traverse the file system, upload and download small files, and execute commands
- **tasks**: Query the status and results of asynchronous operations (most endpoints are asynchronous, and will run in the background until complete.)
- **reservations**: Make and amend requests for NERSC compute resources ahead of time (Currently under implementation)

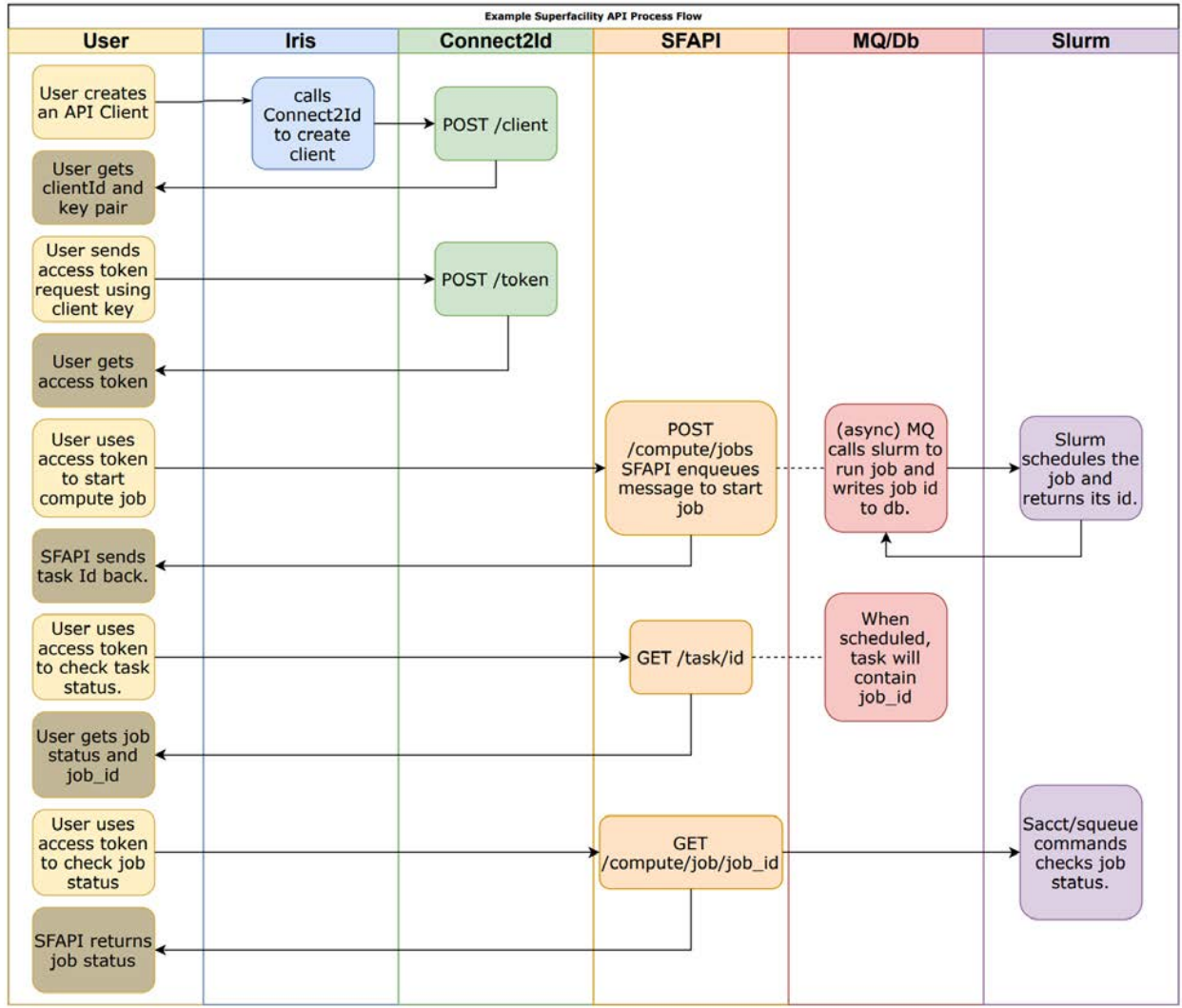

Figure 14. Graph call of a typical scientific workflow, using the NERSC API.

The NERSC API is designed for high throughput, concurrent access. For this reason, it is deployed on our in-house cloud, Spin. It runs as a set of scalable, containerized microservices so the maintenance team can monitor its performance and easily scale up each component in case of unforeseen demand.

It is essential that use of the API be secure, as well as easy to use. We conducted a full security review of the API as a final stage of its development to ensure its authentication and access control model and internal mechanisms for integrating with backend NERSC systems were consistent with security policy and practice. A particular challenge in this area is that NERSC requires the use of MFA, typically by interactively entering a short-lived one-time password (OTP); however, the API is designed to be used non-interactively. To satisfy both requirements, we devised a scheme where users obtain a long-lived OAuth 2.0 credential after successful MFA. To mitigate the risk of the credential being stolen and used elsewhere, it is locked to a small IP address range known only to the user. There are currently two modes of access a user can choose when creating an API credential:
1. Read-only access. Any user can generate credentials with this level of access without additional review; each credential has a 6-month lifetime.
2. Read/Write/Execute access. Users must first request permission to create credentials with this level of access; their application is reviewed by NERSC staff and granted if there is suitable scientific merit and security controls on the system(s) where credentials will be used. Credentials of this type have a more limited duration.

**Gaps/Next Steps**

The NERSC API is still missing the ability for users to programmatically request reservations. In the near future, we're designing a semi-automated workflow for users to request, manage and cancel reservations. The system should be able to handle overlapping reservation intervals, cancellations and notifications to all parties involved.

Besides reservations, we plan to support better logging and monitoring around the API. Once better monitoring is in place, usage statistics should give us insight into ways to improve the API, add higher level functionality, etc. We also continue to review and expand the API end points, based on the needs of our science partners.

Finally, with the Superfacility API officially launched, NERSC will begin a more concerned outreach and training program.

## 3.10: Spin

> **Science teams supported**: ALS, DESC, DESI, JGI, LCLS, LZ, NCEM

As data-driven research has evolved, the need for *edge services* – network resources that sit at the edge of the HPC environment and help with workflow orchestration, results tracking, and the organization and dissemination of results – has emerged. For example:
- Web sites (or *science gateways*) enable researchers to easily view and steer workflow progress and end-users to access HPC storage to download results
- Persistent databases that are available across thousands of HPC job runs allow parameters and/or intermediate results to be easily cataloged and tracked
- Callable API services allow HPC jobs to quickly query data items, exercise ML inference, or perform other tasks at high transaction rates

Edge services can take many forms, but they are increasingly integral to data-driven science, improving efficiency, functionality, and accessibility. Launched in May 2018, Spin is an on-premise container-based platform where users can build and deploy their own edge services on secure, managed infrastructure adjacent to NERSC HPC systems and storage.

Over the course of the Superfacility project, numerous training courses and programs were offered to help users get started with Spin for their science projects, including many of the project engagements described above. Spin was also re-launched in April 2020 on the Kubernetes-based Rancher 2 platform, and new cluster nodes with 4TB of RAM were brought online in August 2020 to support additional use cases with large memory footprints.

As of the writing of this report, interest in Spin is substantial, and continues to grow:
- 18 SpinUp workshops were held for a total of 228 workshop attendees
- 35 Spin office hours sessions were held
- 57 NERSC projects have requested access to Spin
- 500+ edge microservices have been deployed

Superfacility project engagements have been encouraged and supported in their use of Spin to help NERSC staff understand a breadth of edge services needs, resulting in a diverse set of deployments, as follows:
- ALS is using Spin to host their user data portal
- DESC is operating workflow management tools built on the Modern Research Data Portal framework
- DESI has built a suite of web services to help manage data and monitor workflows
- JGI is operating Phytozome, an aggregation of plant genomics services built on several science gateways and databases, among others
- KSTAR shares cyclotron-produced data sets with collaborators using an interactive web-based dashboard
- LCLS uses Spin to coordinate their data movement, management and analysis during experiment data-taking
- LZ has developed a web-based viewer to allow collaborators to easily analyze and inspect notable detector events, a browser which contains runs, configurations, and processing status, and the monitoring services for data movement and data processing tools.

## Gaps/Next Steps

As interest in Spin grows and the underlying Kubernetes subsystem becomes more popular within our community, we are seeing not only increased overall usage and demand for scale, but also new use cases that involve a computational component. Increased adoption also highlights the need for advanced security. We hope to address these emerging requirements with new capabilities in several areas:

- Management
  - Improved resource metrics and visualization
  - Resource caps to prevent "noisy neighbor" impact on other users' workloads
- Security
  - Container image scanning to detect software vulnerabilities *before* runtime
  - Automated policy enforcement to patrol for improperly exposed services
- Computational workload support

- Pre-authorizing Spin services to securely submit HPC jobs
- Allocating an HPC partition as a Spin workload cluster to provide access to scratch storage, accelerators, and other HPC node resources
- Integrating batch scheduling with Spin to enable on-the-fly, short-lived virtual clusters to run workloads designed for Kubernetes

## 3.11: Self-managed facilities

As we increasingly support complex user workflows, our systems are becoming more complicated and difficult to tune, which makes it hard to deliver the best performance to the users. To address this we need to understand how we can automate our systems through informed, data-driven approaches. Self-managed facilities is a complementary effort to the Superfacility project designed to identify opportunities and challenges to automating the operation and improving performance of a supercomputing facility through a data-driven approach. This effort consisted of regular meetings to bring in ideas from across the Computing Sciences organization, including NERSC, CRD, and ESnet.

Coinciding with the start of self-managed systems was an effort by NERSC to begin capturing comprehensive system data with the intent of improving system performance through retrospective analysis. This was done by installing lightweight distributed metric service (LDMS) on NERSC systems. NERSC maintains one of the largest data sets of LDMS data of the Department of Energy HPC systems. This has enabled us to perform a variety of studies and tune system performance in ways that were not possible on prior systems. An example of this is tuning adaptive routing that increased performance of applications by 10% [7]. LDMS collects data of network switches, NICs, memory, Lustre and filesystems, and CPU utilization across thousands of counters, thousands of nodes, and thousands of hardware components.

While LDMS data is one of the newest data sets NERSC has incorporated, there are many other data sets that have provided other sources of information. These include user account and job run information (Slurm), operations data, power, temperature (OMNI), and filesystem I/O profiling (Darshan). Efforts have been made by a variety of NERSC staff to incorporate these disparate data sources to create single data sets that we can analyze at a given point in time. This includes incorporating visualization capabilities from OTG (e.g., Grafana), I/O analysis (e.g.,Tokio), and making the LDMS output easier to ingest (CSV to Parquet). NERSC staff benefit from the opportunities to communicate their contributions within the self-managed facilities community via this group and provide focused solutions that amplify each other's efforts. NERSC staff and summer students completed a number of studies in the scope of self-managed systems during the Superfacility project [7-12].

In addition to using systems data to tune our existing systems, having access to comprehensive system data is creating new opportunities for system design. As we map the growth of computation, communication, power, and I/O to inform our next-generation system designs, we want to ensure that Superfacility use cases of the future have the resources they need.

**Gaps/Next Steps**

Many lessons have been learned from the efforts in self-managed systems over the last several years. Specifically we've begun to understand the value of detailed, system-wide data collection, the algorithmic challenges and hardware required to perform analysis of data at these scales, and the gaps between offline and online analysis. These challenges will continue to become more pronounced in future systems as heterogeneity and complexity increase. This motivates the application of scalable modeling techniques and perhaps machine learning that can be quickly deployed as the system continues to evolve in a production environment.

## 3.12: Software-defined networking

> **Science teams supported**: ALS, KSTAR, NCEM

The goal of the software-defined networking (SDN) technology area is to enable seamless data transfer mechanisms for experimental facilities to stream data into Cori and Perlmutter and provide an optimized network path to critical resources like data transfer nodes (DTNs). Software control of the network should allow new flexibility in how to plan the logistics of data, particularly in regard to bandwidth.

Superfacility-related SDN work includes:
1. The NERSC internal network has been extended to NCEM, providing a 4x100G path for direct data transfer from the detector into a compute node.
2. SDN gateways on Cori provide public IPs for data transfers to work.
3. The ability to allocate load-balanced compute nodes on the ARIES interconnect fabric was developed for the project to optimize data transfer paths to the compute nodes.

To support the large data volumes coming from LCLS to NERSC, we improved connectivity between ESnet and SLAC from 1x100Gbps to 2x100Gbps and provided path resiliency. This diverse 100Gbps path between SLAC and ESnet (@Sacramento) is currently used as a backup to the primary 100Gbps connection between SLAC and ESnet (@Sunnyvale). We provided SLAC a method by which to select this secondary path for ExaFEL traffic using the ESnet OSCARS[20] circuit and implementing NERSC-internal routing policy changes. This freed up the primary connection and avoided congestion. In coordination with NERSC, configurations were also added to the SLAC border routers to enable traffic steering from/to the SLAC site.

For Perlmutter deployment we have deployed a new network core which is capable of supporting multiple 400Gbe circuits. In December 2021, we deployed 1.6 Tbps total bandwidth to Perlmutter to support direct data transfers from our peer sites.

In Summer 2021, we upgraded one NERSC border router to be 400Gbe capable; the second router upgrade is scheduled to be completed by mid-2022. We are currently working with ESnet to get our first 400Gbe circuit. Moving to multiple 400Gbe connections to ESnet will help us to provide guaranteed bandwidth reservation for superficiality science teams by logically carving out portions of our ESnet bandwidth for those projects.

---

[20] https://www.es.net/engineering-services/oscars/

**Gaps/Next Steps**

Perlmutter has an Ethernet-compatible HPC fabric, which opens up a lot of opportunities for network innovations. We are working on formalizing the technical details to implement SDN access policies to give a compute job the ability to initiate direct data transfers to the system. This involves working with the supercomputer vendor to deliver some features in future software releases and development work on the NERSC side to integrate the workflow.

Another future functionality is inband network telemetry. We are currently working on deploying a new CFS network fabric. To optimize fabric performance, we want to create an analytics framework based on telemetry data and use the data to program network queues and traffic class specifications.

## 3.13: SENSE

**Science teams supported**: DESC, LCLS

SENSE (SDN for End-to-End Networking @ Exascale) is an ESnet orchestration and intelligence system that provides networked services to domain science workflows. These services can span multiple domains/sites and be presented to different workflows in a highly customized manner. A key objective of the SENSE system is to allow the workflows to access the network in a manner which best facilitates their objectives. SENSE services include Layer 2 point to point network connections, Layer 2 multipoint network topologies, and Layer 3 virtual private network (VPN) services.

The SENSE system also provides a variety of interactive services that allow application workflows to ask open-ended questions about capabilities, negotiate with the networked infrastructure, or request network services in a highly abstract and workflow centric manner. The SENSE services are referred to as "networked services" because in addition to the network elements, SENSE can orchestrate across the elements that connect to the network, such as DTNs, instrument servers, and edge router configurations. Additional information regarding the SENSE system is available in [13,14].

**Superfacility workflows and integration and enhancement of SENSE services**
This activity focused on the LCLS workflow, with the goal of prototyping an end-to-end orchestration framework to enable reliable and predictable data transfer behavior between SLAC and NERSC over ESnet. This is to support time-sensitive services such as deadline scheduling for large data movements, as well as fast feedback for adjustment of LCLS experiments using NERSC compute resources. The LCLS workflow was modified to use the SENSE application programming interface (API) for reserving network bandwidth between SLAC and NERSC and integrating it with the LCLS data movers. For technical details and description of the enhanced workflow operations, please see [13,14].

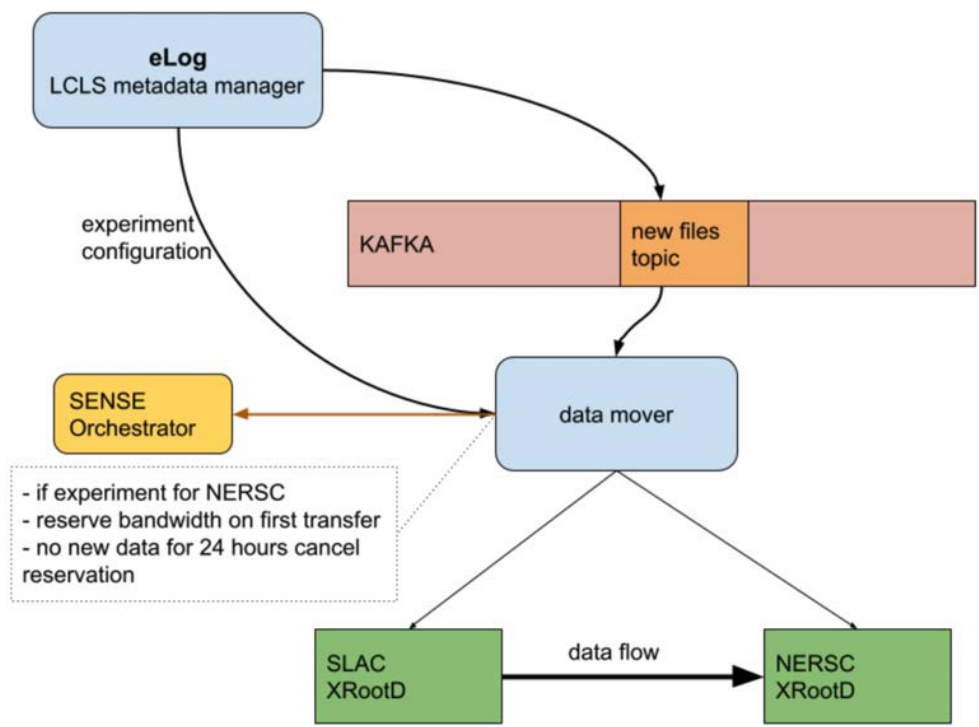

Figure 15. SLAC to NERSC data movement workflow, adapted from [14].

It is expected that the methods used by the LCLS workflow will apply to a broad range of domain science workflows. Work is under way to help others use the SENSE API in a similar manner; this includes working with the Large Hadron Collider (LHC) community to integrate SENSE services into their workflows and data transfer functions based on Rucio/FTS and XRootD. SENSE networked services could be a standard service made available to future Superfacility users. There is also future work to determine how the SENSE API and Superfacility API should be used in coordination for workflow operations.

## Gaps/Next Steps

This system was deployed as a prototype for data transfers between SLAC and NERSC across ESnet [14] for LCLS workflow. There are plans to continue this work to create a persistent system that will allow continued development of this functionality. Future plans include integrating the monitoring system as an automated SENSE service. The end-to-end monitoring system is focused on helping with multi-domain data transfers as part of science workflows. The purpose of this developmental activity is to look at a single end-to-end flow in near-real-time to facilitate understanding of current performance and discovery of issues that may be limiting expected end-to-end throughput. In this context, "end-to-end" covers the network elements along the path and the network stacks inside the end systems. This work was narrowly focused on establishing near-real-time, end-to-end situational awareness with respect to a single flow. As a result, it is not concerned with collection or storage of general data about specific network elements or end systems. This monitoring system is envisioned to be complementary to existing monitoring systems that are focused on characterization of entire infrastructures.

## 3.14: Enabling collaborations to move and share data

> **Science teams supported**: ALS, DESC, DESI, JGI, KSTAR, LCLS, LZ, NCEM

Data migration into and out of NERSC is of critical importance to superfacilities. NERSC offers Globus as a solution for those sites that don't have their own custom data movement software. This offers sites an easy way to get parallel, large-scale, resilient data transfers between institutions. Globus also serves as an excellent way to move data between different file systems at NERSC. In 2020, scientists at NERSC moved 37 PB of data with Globus.

One long-standing issue with Globus is that it doesn't support NERSC's collaboration accounts, which are special internal accounts that permit a many-to-one mapping within a group to facilitate data sharing and shared software installation. Groups that wanted to use collaboration accounts with Globus would have to transfer the data as themselves, then ask NERSC staff to do a "chown" to the collaboration account, which is a laborious process. As part of this project, special Globus endpoints were deployed that allowed NERSC scientists to write data directly to NERSC as their collaboration account. This has been an enormous time saver for these groups and greatly facilitated data movement at NERSC.

Growing data volumes have made new challenges for scientists at NERSC. For many years, NERSC has provided http pages for data sharing, but this has become increasingly untenable as shared volumes move into the TB-PB range. For the short term, we have deployed the capability for scientists to share data via Globus Shared Collections. In 2020, scientists used these endpoints to share 7.5 PB of data. As these numbers grow, we are also gaining new users with less experience in transferring large files efficiently. We are therefore working on a "PB Data Portal" to facilitate sharing these large volumes of scientific data.

**Gaps/Next Steps**

Movement between file systems at NERSC is becoming more challenging as data volumes grow. In order to offer the best performance at the optimal cost, we have a tiered file system at NERSC. The top of the tier is a high-performance Lustre scratch file system with limited capacity, next is a moderate-performance Spectrum Scale file system with enough capacity to handle multiple years of NERSC data, and finally there is an HPSS archive tape system that houses all NERSC data since NERSC's inception. Users must manually move their data between tiers, which demands a substantial amount of time and effort. To make this easier, we have been investigating tools to more easily move data between tiers. We deployed a test instance of GHI, a software package that allows users to use the Spectrum Scale file system as a front end for the HPSS tape archive. We tested GHI with several Superfacility partners and it was found to satisfy most of the user requirements. NERSC is currently considering the hardware and personnel costs needed for deployment. We have also engaged SchedMD to develop capability to stage data between tiers as part of job submission. Future work will continue to explore these and other tools for data movement.

## 3.15: Managing data within NERSC: Data Dashboard and PI Toolbox

**Science teams supported**: ALS, DESC, DESI, JGI, LCLS, LZ

Wrangling data at a scientific computing center can be a major challenge for users, particularly when quotas may impact their ability to utilize resources. In such an environment, a task as simple as listing space usage for one's files can take hours. NERSC has roughly 60 PBs of shared storage utilizing more than 3.5 billion files and directories, and a 250 PB high-performance tape archive, all accessible from the Cori and Perlmutter supercomputers. As data volumes increase exponentially, managing data is becoming a larger burden on scientists. To ease the pain, we designed and built a "Data Dashboard" (Figure 15). Here, in a web-enabled visual application, our 8,000+ users can easily review their usage against quotas, discover patterns, and identify candidate files for archiving or deletion. The Data Dashboard has been deployed as part of the MyNERSC user web site.

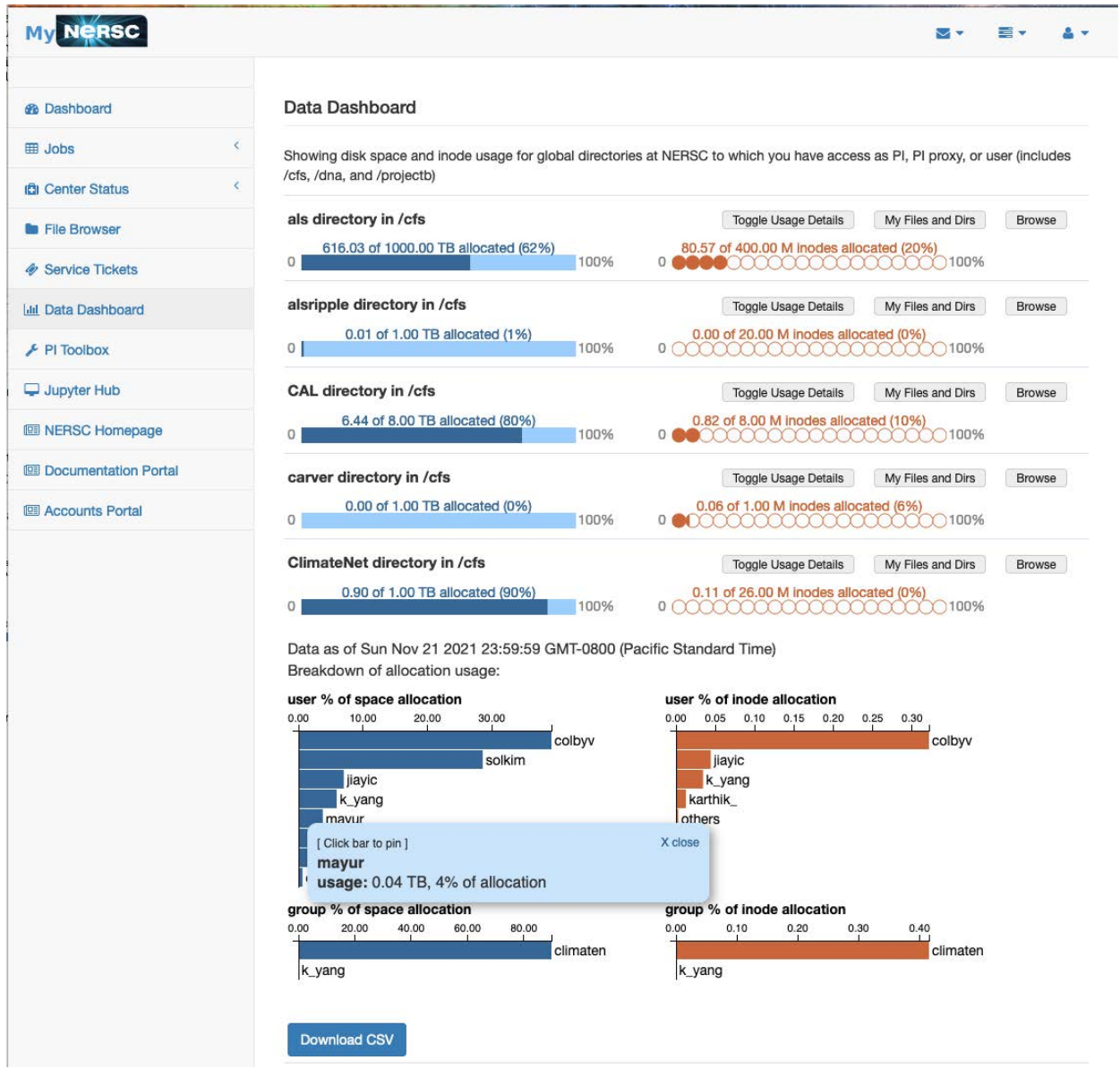

*Figure 15. The Data Dashboard shows space and inode usage information for each user on a per-project basis.*

Another source of pain for HPC users is the drift of file permissions away from settings that facilitate collaboration within research teams. As members come and go, the principal investigator may need to reassign permissions and ownership of shared files. Typically, making such changes requires filing a support ticket and taking the time of a consultant to make the actual changes. PIs have long wanted to be able to make such changes themselves, so we have developed a "PI Toolbox" (Figure 16) to allow them to directly control the permissions of their files and directories. Deployed as a separate tab within MyNERSC, the toolbox is now fully operational. It allows users to browse their project's files on the Community File System, change the group or permissions of any files or directories, or simply make the entire shared project directory group readable.

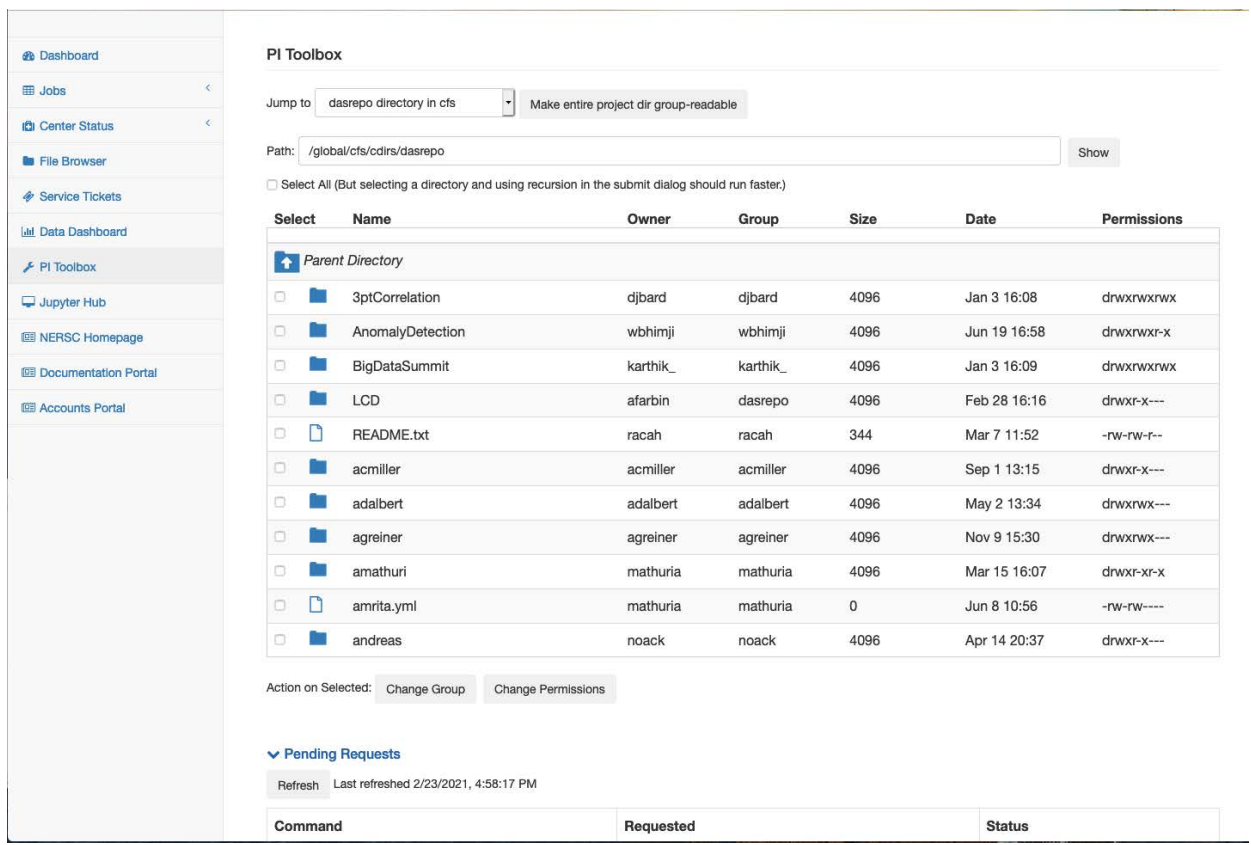

*Figure 16. The PI Toolbox allows project leaders to update permissions on files and directories in shared data storage.*

Developing these tools involved building a new, generalizable framework for data management. Existing commercial options lack the access management we need to release to many users, and noncommercial solutions have thus far been customized to narrow user segments or specific storage systems. We therefore built a new framework on tools that come with common file system software, like Spectrum Scale, Lustre, and HPSS.

Both the Data Dashboard and the PI Toolbox leverage web APIs to accomplish their aims. As web-based applications, they serve as proofs-of-concept for the API-based interaction between users and the resources of a Superfacility. Our usage statistics bear out the level of user interest in such tools. In calendar year 2021, 1,574 users took advantage of the dashboard's capabilities and 1,326 accessed the PI toolbox, including users from all 8 of our Superfacility science engagements.

## Gaps / Next Steps

Because the Superfacility API's authentication model does not yet support applications acting on behalf of arbitrary users, we decided to initially build our tools with the existing NEWT API. We have been working closely with developers of the Superfacility API to ensure that all the functionality we need is eventually enabled in the latter. We plan to port the system to use the Superfacility API in the future.

We are working to add an intelligent archiver application that suggests files and directories to be archived and enables the user to easily initiate archiving operations. Users will be able to customize the suggestion criteria to match their own workflow. Aside from the challenge of creating a sufficiently flexible tool, we will need to be particularly careful how we design the tool's behaviors, since NERSC users have such a wide variety of workflows. We will also need to develop smart algorithms to ensure that archiving suggestions are appropriate and never annoying. We will continue to add functionality to the PI Toolbox, such as more one-click permissions options and the ability to change ownership of top-level project directories. The new features will be driven, as always, by user requirements.

## 3.16: HDF5

**Science teams supported**: DESC, LCLS, NCEM

HDF5 is a general-purpose storage middleware used in many scientific applications and supported by many analytical tools. HDF5 provides a rich data model that allows representing very complex data objects with a wide variety of metadata, a high-performance software library that implements that data model, and a self-describing, portable file format that has no limit on the number or size of data objects in the collection. As a result, HDF5 has been used by many science applications to store and share their data. In the Superfacility project, LCLS and NCEM use HDF5 to store and share data products using the HDF5 file format. Toward supporting HDF5 requirements in the Superfacility project, we have worked on providing efficient access to data and metadata stored in files, understanding data usage for supporting optimizations, and mapping LCLS's XTC2 data format and HDF5. These activities have been partially funded by other ASCR-funded projects, including the "EOD-HDF5: Advancing HDF5 to support experimental and observational data" and the "LLANA: LCLS-LBNL data Analytics collaboration" projects.

Systematic capture of extensive, useful science metadata and provenance requires an easy-to-use strategy to automatically record information throughout the data life cycle, without posing significant performance overhead. Toward that goal, we have developed a virtual object layer (VOL) connector for HDF5. The VOL connector, called H5Prov, transparently intercepts HDF5 calls and records operations at multiple levels, namely file, group, data set, and data element levels.

Scientific applications often store data sets in self-describing data file formats, such as HDF5 and netCDF. However, efficient search of the metadata within these files remains challenging due to the sheer size of the datasets. Existing solutions extract the metadata and store it in external database management systems (DBMS) to locate desired data. This practice introduces significant overhead and complexity in extraction and querying. We developed a novel metadata indexing and querying service (MIQS) for self-describing formats that removes the external DBMS and utilizes in-memory index to achieve efficient metadata searching. Our evaluation of searching metadata stored in 100 HDF5 files from the Baryon Oscillation Spectroscopic Survey (BOSS, a precursor to the DESI experiment) demonstrated that MIQS achieved up to 99% time reduction in index construction compared to MongoDB and up to 172,000 times higher throughput than MongoDB in searching their respective indexes (as shown in Figure 17), while reducing in memory footprint by 75%.

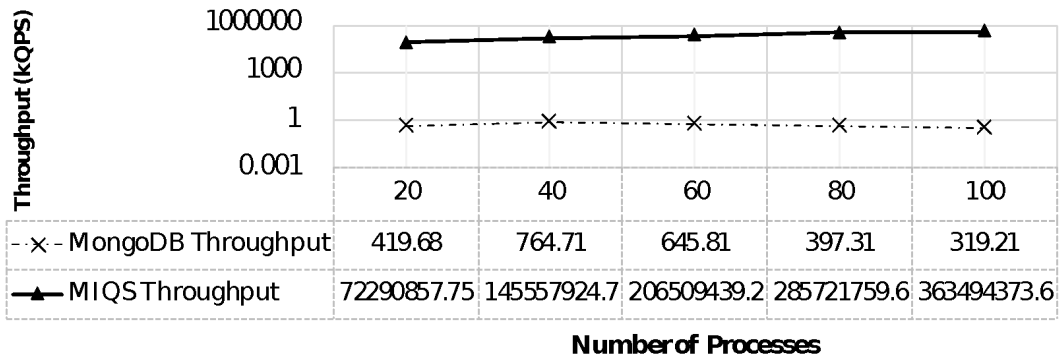

Figure 17. Query performance (throughput in x1000 queries per second) of MIQS compared with MongoDB. Dataset used: Baryon Oscillation Spectroscopic Survey (BOSS) data stored in 100 HDF5 files with 144 million metadata attributes describing 1.5 million data objects. In this query, we compared the performance of querying 16 attributes. Latency for searching these attributes with MongoDB was 5 min using 100 processes, compared to 0.29 ms with MIQS.

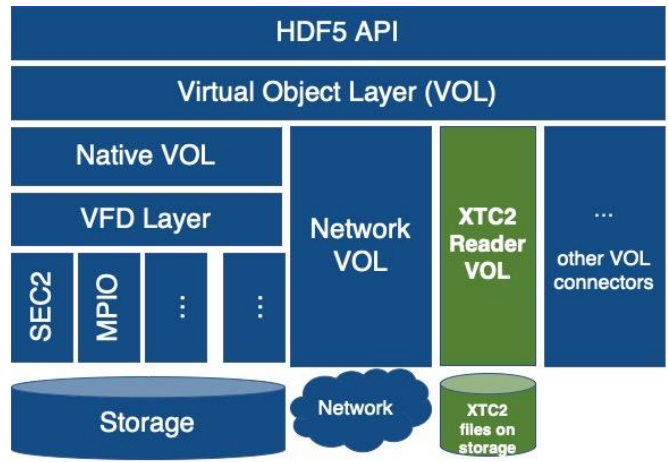

Figure 18. HDF5 virtual object layer (VOL) connector to read XTC2 formatted files.

XTC2 is a scientific data format used in LCLS data analysis codes, such as *psana*, and contains data produced by the online data acquisition system. As a part of efforts on the SLAC-NERSC Superfacility project, we developed a VOL connector for reading XTC2 data from the HDF5 API to allow HDF5 applications and visualizations to read XTC2 files. As shown in Figure 18, the XTC2 VOL connector allows HDF5 applications to use XTC2 formatted files directly, without requiring to change the code. In the initial prototype, the VOL connector was able to solve simple data structure mapping between XTC2 and HDF5.[21] [22] We have also observed that the XTC2 format is implemented in a compact way, where raw data is encapsulated with multiple layers of abstractions, and each layer uses different iteration methods. Some data structures in XTC2 have a metadata structure for indexing that makes a direct HDF5 mapping for these XTC2 objects difficult and results in poor performance. Further tuning of the current implementation is needed to optimize this mapping.

## Gaps/Next Steps

While HDF5 is used heavily by a large number of applications in the experimental and observational science (EOS) facilities, several features need further improvements. Here are a few features that have been requested by EOS facilities that are applicable in a Superfacility environment.

**Native support for sparse data management**: Working with sparse matrices and other sparsely populated data structures is a common need in many science, engineering, and mathematical domains, but it is frequently overlooked and underserved by data management frameworks and storage packages. Researchers from NCEM have been using workarounds for storing their sparse data by converting to a coordinate-based format and recovering it when analysis applications read the data. LCLS and Fermi applications also have asked for native support for sparse data in HDF5.

**Improved performance for managing variable-length data**: Similar to sparse data, data structures in experimental and observational data are often of variable length. HDF5 currently has API functionality for supporting variable-length data; however, the performance of storing and reading this type of data is poor. Our initial studies showed better data structures that can improve performance but need further R&D to integrate them into HDF5.

**Improve performance of streamed access to dataset records**: Another inherent characteristic of Superfacility is streaming data, where data records are often appended to existing data. This is a common case in High Energy Physics, where events are processed and added to existing data records. With some initial work, we have shown up to 10x improvement with optimized append operations. These improvements have to be tested further with real use cases and integrated into HDF5.

**Support for seamless I/O of data containers stored across multiple and heterogeneous file / data management systems**: In a Superfacility environment, a single file could be distributed across multiple file systems within a facility (scratch, community file system, HPSS, etc.) and across multiple facilities (e.g., NERSC, ALCF, and OLCF), and multiple environments (HPC, cloud, edge, etc.). HDF5 has the baseline concepts of data containers, where it currently

---

[21] https://github.com/slac-lcls/lcls2/tree/dev_tony/xtc_vol
[22] https://drive.google.com/file/d/1C-YMYgStIA5_fo51z3RWp_ma3cz5FT2l

supports files stored within a single file system. The HDF5 container (file) could be split across various file systems and environments. This allows grouping all relevant data objects and their corresponding metadata within a single container. In this scenario, seamless access to data that is located in different storage systems provides flexibility to users to view data containers as single entities, bringing a true "virtual data facility" idea into reality. Designing and developing this concept in HDF5 requires solving challenges related to authentication across multiple models of storage, efficient data transfer methods, non-uniform data access times, and more. Advanced methods for understanding data accesses to cache, prefetch, and creating materialized views of data that users can access with high performance are also needed.

# 4: Science Impact

We met our project goal to have at least three of our science application engagements demonstrate automated pipelines that analyze data from remote facilities at large scale, without routine human intervention, using these capabilities:
- Near-real-time computing support
- Dynamic, high-performance networking
- Data management and movement tools, including Globus
- API-driven automation
- HPC-scale notebooks via Jupyter
- Authentication using Federated Identity
- Container-based edge services supported via Spin.

In several cases (LZ, DESI, LCLS, and NCEM), we have gone beyond demonstrations and can now provide production-level services for their experiment teams.

| Science engagement | Pipeline | Features used to meet goal |
|---|---|---|
| ALS | Automated data movement and analysis | ● **Near-real-time computing** via the real-time QOS<br>● **Dynamic, high-performance networking**, e.g., one beamline pushes data automatically to CFS using Globus over LBLnet/SciDMZ<br>● **Data management and movement tools** in several areas, including the Share app ([alsshare.lbl.gov](alsshare.lbl.gov)), Catalog app, Mover app, and Tiled.<br>● **API-driven automation** via their own APIs, but also using the Globus and SF API.<br>● **HPC-scale notebooks via Jupyter**<br>● **Authentication using Federated Identity** and Orcid |

| | | • **Container-based edge services via Spin**, used to support the data management tools |
|---|---|---|
| DESI | Automated nightly data movement from telescope to NERSC, and deadline-driven data analysis with telescope operations starting in 2020. | • **Real-time computing support**, using the real-time queue<br>• **Dynamic, high-performance networking**, using rsync to automatically transfer data from Kitt Peak to NERSC CFS<br>• **Data management and movement tools**, mostly homegrown series of cronjobs and monitor tasks |
| LCLS | Automated data movement and analysis from several experiments running at high data rate end stations during 2020 and 2021. | • **Real-time computing support** used for running experiments via the real-time QOS and reservations<br>• **Dynamic, high-performance networking** via a high-speed connection on ESNet reserved via the SENSE API<br>• **Data management and movement tools**, using an XRootD-based solution<br>• **API-driven automation**, using the LCLS API to integrate experiment status with the data processing pipeline at NERSC (e.g. completed run number, and status of data movement)<br>• **Container-based edge services via Spin**, used to orchestrate workflows |
| LZ | Automated 24/7 data analysis from the dark matter detector, with commissioning starting in June 2021. | • **Real-time computing support** via the real-time queue for continuous data analysis of the running experiment<br>• **Data management and movement tools** used to automate data movement between experiment site, NERSC and UK data center, via ESnet<br>• **Container-based edge services via Spin** used to coordinate data movement and compute job orchestration |
| NCEM | Automated workflow pulling data from the 4D STEM camera to Cori for near-real-time data processing, starting in late 2021. | • **Real-time computing support**, via the debug and real-time QOS<br>• **Dynamic, high-performance networking**, with the 4D stem camera sitting on an extension of NERSC's network and using SDN to pull ~40GB/sec onto Cori and Perlmutter compute nodes<br>• **API-driven automation** via the Distiller app in Spin which monitors their microscope data and submits jobs via the SF API - a "one |

|  |  | click" solution |
|---|---|---|

In this section we describe in more detail how the Superfacility Project work has impacted the work of our science engagements.

## 4.1: ALS

> **Key Superfacility needs:** NESAP, Policies, Jupyter, Scheduling, Resiliency, Federated ID, API, Spin, Self-managed Systems, Data movement, Data management.

The ALS has deployed a number of development and production projects in Spin. First, a data portal[23] was deployed alongside databases, app server, and several other services supporting workflows for ingesting ALS data. For example, these services access tomography data that is moved from the ALS microtomography beamline to NERSC's community file system (CFS). This data movement service, in turn, leverages improvement to NERSC Globus infrastructure that allowed for writing to NERSC's file systems using a collaboration (i.e. "machine") account. Once the data lands at CFS, the ALS user can simply search and browse their data in the portal based on the metadata.

Second, a service was launched[24] that streamlines data sharing based NERSC's Globus share endpoint and integrates with ALS's userportal ALSHub. When a beamline scientist creates a Globus share for an experiment, this service automatically populates the share with the collaborators of that experiment pulling the right data from ALSHub. Third, a project was created where ALS collaborates with BNL and ANL called "AI/ML for Multi-Modal (AIMM)" that supports data access and data labeling/tagging services on Spin.

**Future plans**
With some beamlines at the ALS now automatically transferring data sets to NERSC as they are collected, an upcoming development will be to enable an ALS Share directory to be automatically set up in advance of data collection, and new data sets will be routed to the that directory so a user and the assigned collaborators can access the data very soon after it is collected. Furthermore, ALS was a key engagement to develop the functionality of the Superfacility API and is currently incorporating the API into their services to kick off standardized computing jobs (or other workloads) for data that has reached NERSC file systems. Finally, ALS envisions all of its users to be able to repeat at NERSC the same analysis that they used during an experiment. ALS intends to use customized Jupyter notebooks and even a customized JupyterHub for that purpose.

---

[23] https://dataportal.als.lbl.gov/static/user-login.html
[24] http://alsshare.lbl.gov

## 4.2: DESC

> **Key Superfacility needs:** NESAP, Policies, Jupyter, Resiliency, Federated ID, API, Spin, SDN, Data movement, Data management.

DESC has one of the largest compute and storage allocations at NERSC, which is used primarily to produce high-fidelity simulations of the sky and mock telescope images. They are part of the NESAP for Data program, which is helping them port their simulation code to GPUs and optimise. The simulations and initial stages of data processing for their second data challenge was concluded last year. The resulting dataset has been publicly shared via a Modern Research Data Portal,[25] a Globus-based infrastructure for sharing data to the wider community.[26] DESC is currently developing workflow systems, using Spin, to run its image reprocessing and analysis pipelines for those data and eventually for real data from the Rubin Observatory.

DESC is the project with perhaps the largest numbers of users at NERSC, with more than 450 users from around the world. As a result they are in dire need of tools to manage and monitor their own users. They have been a driving force behind the PI and Data Dashboards, which help them handle administrative operations like `chown` themselves without having to file a ticket for NERSC staff every time. They are also a key use case for automated user and data management via the API. As they work to develop their analysis pipelines, the API will provide much of the functionality they need for data movement and job submission.

DESC uses Jupyter notebooks as their main analysis tool. DESC staff create curated Jupyter environments that contain custom software packages, and notebooks are shared across the entire collaboration. Since their analysis work includes large amounts of data processing, they provide a key motivating use case for the work to support back-end parallelization of code using dask.

**Future plans**
NERSC will be the primary host for that data, which will increase in size and complexity over time. DESC is preparing for the first data from Rubin in 2024. The collaboration is using NERSC as their primary facility for the development of analysis pipelines and targeted reprocessing to evaluate and characterize possible systematics. Some of the upcoming tasks will involve the generation of synthetic data, and NERSC will be one of the major computational centers for facilitating this work.

---

[25] https://data.lsstdesc.org/

[26] https://www.nersc.gov/news-publications/nersc-news/science-news/2021/nersc-and-esnet-take-scientific-data-collaboration-to-the-next-level/

As the collaboration develops and finalizes its analysis pipelines, they will continue to work closely with NERSC to ensure they have support for their simulation production, processing, and end user analysis. A key activity in the next two years will be developing workflows that can operate across the various computing resources available to the collaboration. NERSC's work in cross-facility workflows will help inform that effort.

## 4.3: DESI

> **Key Superfacility needs:** NESAP, Policies, Jupyter, Scheduling, Resiliency, Spin, Self-managed Systems, Data movement, Data management.

To help DESI achieve quick throughput of its nightly exposure data, they perform this processing using NERSC's real-time queue. As of November 2021, they have used about 5.6 million NERSC hours for this processing in the real-time queue.

In addition to the issues related to scientific analysis, as a large collaboration with 200+ members (as of the latest collaboration meeting attendance), DESI faces issues with file permissions, data management, and data transfer. Two improvements within the Superfacility project have been helpful in this regard:

- **Using Globus as a collaboration account**: DESI makes routine use of collaboration accounts, and being able to transfer their data via a collaboration account makes creating and maintaining file ownership and permissions a lot more streamlined.
- **The new [Data Dashboard](#)**: Makes it easier to change permissions and ownership of files via `chown`.

**Future plans**
Through extensive collaboration with the NESAP program, the DESI Spectroscopic pipeline is now largely GPU-enabled. This will enable DESI to transition its production pipeline over to *Perlmutter* in the near future. DESI will continue to send its nightly exposure data to NERSC for near-real-time processing throughout the remaining years of their survey. They will also perform large-scale re-analyses of their data 1-2 times per year. NESAP efforts have enabled this processing to take place on the scale of the whole *Perlmutter* GPU partition in approximately 40 minutes (see Figure 19).

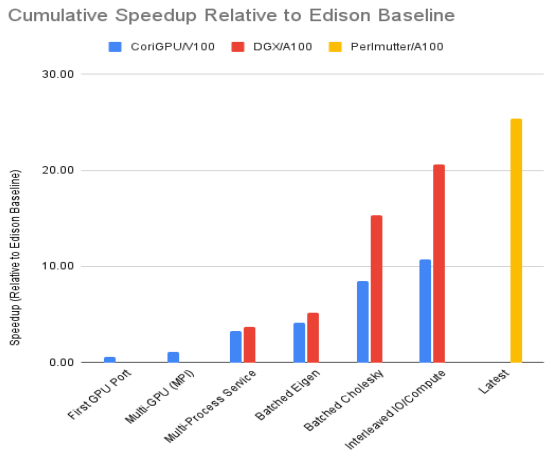 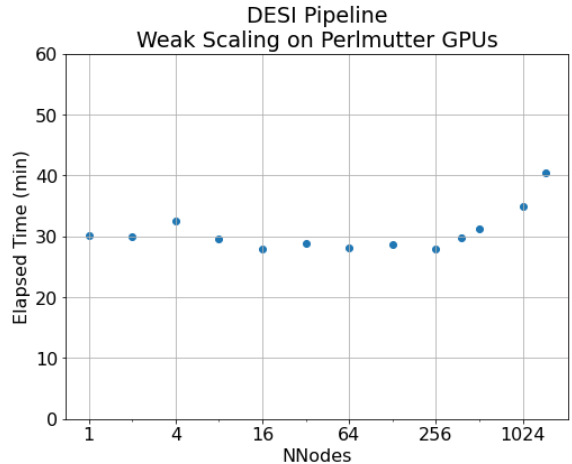

*Figure 19. (L) The speedup achieved throughout the NESAP for Perlmutter effort. (R) The weak scaling of the DESI pipeline, all the way up to the full size of Perlmutter Phase I.*

To continue to enable DESI to make productive use of NERSC systems, DESI requires a robust and stable system that enables them to continue running their scientific pipeline but also provides a productive atmosphere for a large collaboration to perform, share, and curate their work. Since data are central to DESI's analysis, they require reliable access to read/write their data via robust file systems. They have historically used the NERSC workflow nodes to coordinate their data transfer and workflow pipeline, but may eventually transition to Spin/Superfacility API.

## 4.4: JGI

**Key Superfacility needs:** Policies, Resiliency, Federated ID, API, Spin, Data movement, Data management.

The creation of Spin as a service hosting platform has given a convenient location for most JGI services to relocate after the retirement of their initial hosting hardware. These services additionally benefit from the closer connection to NERSC file systems, archives, and computational capacity.

Assembly of the largest and most difficult metagenome datasets benefit from additional resource scheduling options made available by the Superfacility project, via reservations of compute nodes. They are also able to make use of the new capability to run preemptible jobs, as some JGI workflows consist of many small tasks that fit well with preemption.

A large number of JGI users and projects have transitioned to using Globus for their data transfer needs, both between various NERSC resources and to the outside world. The strongest example is the JAWS workflow management tool, which leverages new Globus functionality to automate the transfer of entire workflows and datasets between different computational platforms and institutions.

**Future plans**
JGI is leading the way in developing cross-facility workflows. The long-term strategic plan of JGI is to diversify the number of venues where their computational workload can be run and to automate the distribution of work to those various facilities. The JAWS orchestrator is planned to be the core mechanism of this change, and it is already using a number of the Superfacility project components to interface with NERSC. If the Superfacility model expands beyond NERSC to other DOE facilities, JGI would benefit by streamlining its connections to the JAWS workflow management tool.

## 4.5: KSTAR

> **Key Superfacility needs:** Jupyter, Scheduling, Resiliency, API, Spin, SDN, Data movement.

KSTAR was a relative latecomer to the Superficiality project, joining in late 2020. Since then, NERSC has worked with the KSTAR development team to:
- Assist in application profiling, GPU porting, and preparing for *Perlmutter*
- Assist in using NERSC resources like Shifter and Cray MPICH
- Help set up testing reservations.

On-going work includes:
- Enable Shifter on the DTNs to help reduce software stack wrangling
- Open the *Perlmutter* network to streaming data from the KSTAR DTNs, maybe via SDN or something similar.

**Future plans**
This effort is the beginning of a much larger effort to integrate fusion experiments like KSTAR with HPC resources like NERSC to enable computational tasks that are not possible with on-site computing. The current framework is, to some extent, a pilot effort meant to test some early types of data analysis and machine learning tasks. On the fusion side, larger open questions are related to what kinds of tasks are most useful for streaming to a remote HPC center; on the NERSC side, a key question is what kinds of infrastructure are needed to support these workloads.

KSTAR's future needs at NERSC will be largely related to near-real-time computing. This includes:

- The ability to stream data from external sources in near-real-time
- Near-real-time availability of compute resources. "real-time" requirements do, in fact, mean real time – minutes count. If the jobs were to get stuck in the queue for 5 minutes, they may not finish in time to be used by the KSTAR team.
- Other needs are related to resilient systems. For an experiment to rely on NERSC for near-real-time turnaround, it needs not only compute resources, but also robust queue resources, robust networking resources, and robust database and Spin resources.

## 4.6: LCLS

**Key Superfacility needs:** NESAP, Policies, Jupyter, Scheduling, Resiliency, Federated ID, API, Spin, SDN, Self-managed Systems, SENSE, Data movement, Data management, HDF5.

The Superfacility team (at NERSC) has worked with the AutoSFX developers (at the LCLS) as well as the CCTBX developers (at LBNL) to enable live data processing during LCLS beamtimes. In 2020 this was demonstrated in principle for the AutoSFX toolchain, and it was used for production data analysis in 2020, and 2021 for the CCTBX toolchain [16,17].

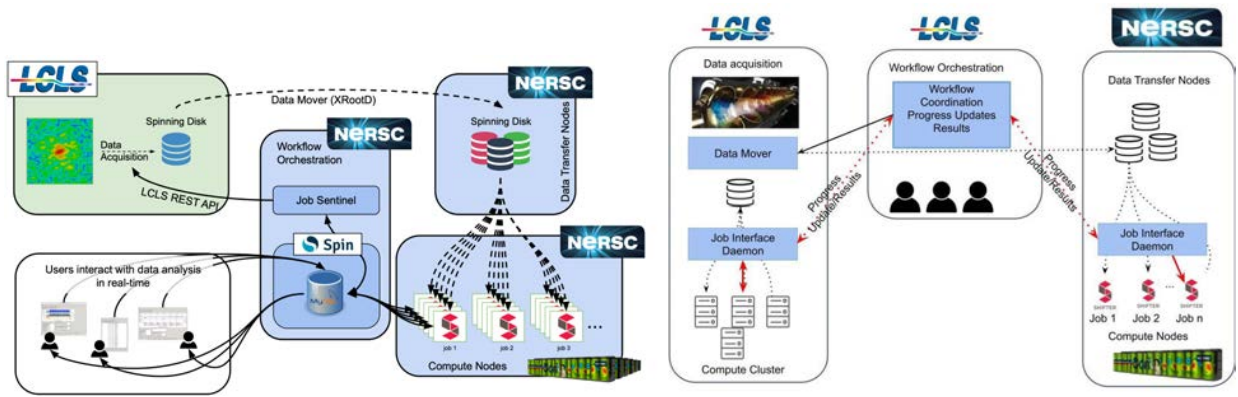

*Figure 20. the Superfacility approach used by the CCTBX (left) and AutoSFX (right) toolchains.*

Figure 20 shows the LCLS workflow:
1. Bandwidth on the ESNet network is reserved ahead of the beamtime via the SENSE project – as well as compute reservations at NERSC
2. Data is acquired at the LCLS and automatically copied to NERSC via ESNet
3. Once at NERSC, a helper application marshals data analysis workers (running in shifter containers on compute nodes)
4. Live data processing results are communicated back to the users (usually working remotely or at the LCLS) in real time.

The main difference between the AutoSFX and the CCTBX workflows is that job marshaling happens either at NERSC (running on a login node and on Spin), exposing results via a REST API, or at the LCLS using the Superfacility API. Both approaches have been shown to be flexible and resilient. The CCTBX production runs during 2020 were able to complete real-time data analysis within as little as 5 minutes of the data being collected (and no longer than 20 minutes). The production runs in 2021 where able to exercise preemptible reservations.

**Future plans**
Work in 2020, and 2021 has shown that the superfacily approach works for them. Future work will therefore revolve around: i) hardening existing workflows to be more reliable (and in particular include fewer human-in-the-loop steps – for example, by making use of Federated ID, and the Compute Reservations SFAPI endpoint, once they are available); expanding their toolchain to include new technologies (e.g., using Perlmutter and GPU-based data analysis, and using preemptible reservations); adding new algorithms to the existing tool chains (e.g., adding SPI and Machine Learning); and incorporating additional ASCR computing facilities (e.g., OCLF, with which the CCTBX, MTIP, and AutoSFX teams have a working relationship).

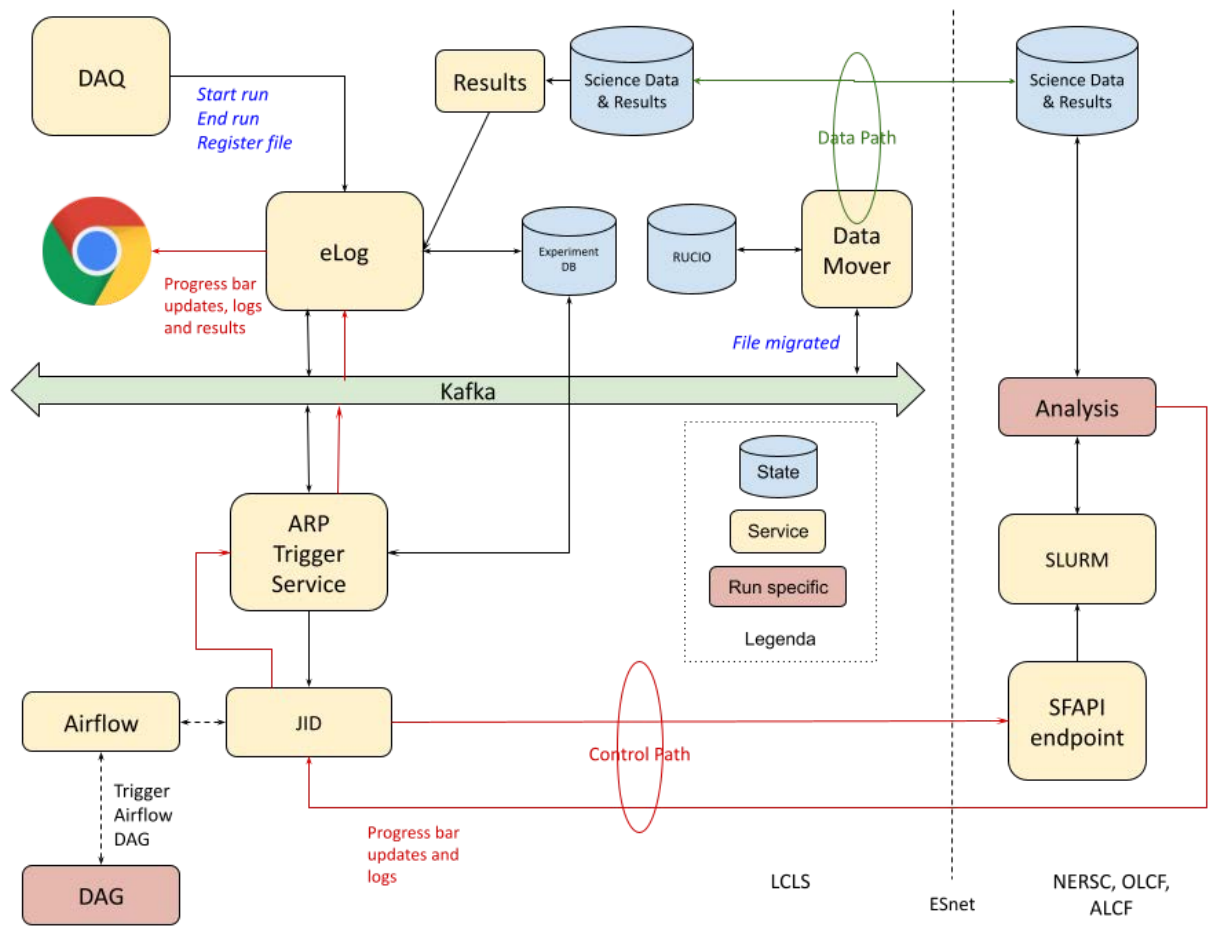

*Figure 21. Role of the SFAPI within the LCLS data analysis framework. Workflow-specific components are customized in the Airflow DAG and in the analysis job itself.*

There is also a general appetite to explore new technologies (e.g., Jupyter and new data management technologies), but no concrete plans to this end have materialized yet.

## 4.7: LZ

> **Key Superfacility needs:** NESAP, Policies, Jupyter, Scheduling, Resiliency, Federated ID, API, Spin, Data movement, Data management.

LZ's workflow takes data from the detector up to the surface facility, then on to NERSC where it is processed. The raw and derived data products are archived and also sent to the UK data center. This full workflow is monitored via dashboards in Spin, which can quickly flag any change in data rates or data quality. NERSC is the primary data archive for LZ, so the team has worked closely with the NERSC team to test new data management tools, such as GHI (see Section 3.14) to enable automated data movement and archiving.

LZ also uses NERSC for simulation production and is part of the NESAP for Data program that is helping port their code to GPUs in preparation for *Perlmutter*. They produce detector simulations annually, with data products available to their collaboration. LZ has approximately 200 members with accounts at NERSC, and they have motivated some of the developments in user management via the API and the PI Dashboard.

Since LZ operates 24/7, they are deeply concerned with the availability of NERSC resources. They have the capability to operate with reduced analysis capability for several days, using compute and storage on site at Sanford, but that leaves them vulnerable to detector problems that may be missed with a lower fidelity analysis. The team has worked closely with NERSC on developing our resiliency plans to keep more parts of NERSC infrastructure available during routine maintenances and power work, and they have also enthusiastically participated in the nascent work to port workflow to other computing sites.

**Future plans**
LZ is designed to operate for five years and will continue to use NERSC as the primary data center for the lifetime of the experiment. LZ is therefore strongly supportive of multi-year allocations. It is hard for an experiment to plan ahead when there is uncertainty about the amount of compute time they will be allocated and where it will be assigned. The NESAP program will continue to support LZ as they develop their simulation code for GPU processors, as *Perlmutter* will be their primary platform for the lifetime of their experiment.

LZ is a small experiment, with limited scientist time available to develop new software infrastructure but with multi-PB-scale datasets. Because of this, LZ (and other "small" experiments) needs to adopt existing, proven tools and technologies that can be easily adapted to their needs [18]. They would benefit from a set of recommended workflow components that can run across sites and are easily maintainable and well documented. This would also be of benefit to all NERSC users and is something the Superfacility team is considering for future work.

## 4.8: NCEM

> **Key Superfacility needs:** Jupyter, Scheduling, Federated ID, API, Spin, SDN, Data movement.

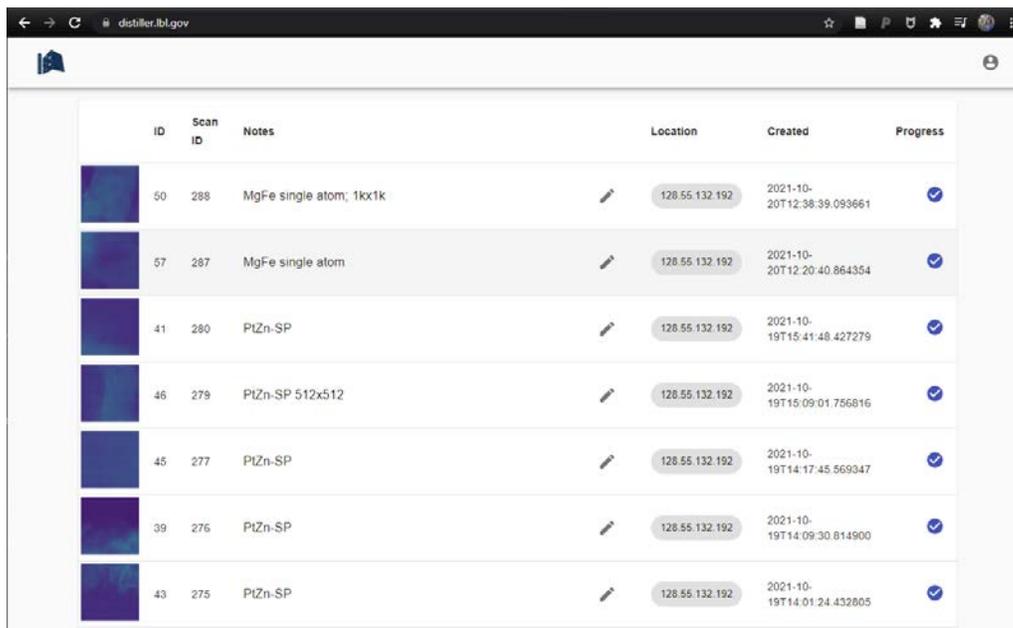

*Figure 22. Screenshot of the NCEM Distiller app, showing the progress of data analysis tasks.*

NCEM's 4D Camera produces so much data that it precluded NCEM from using typical workstations for storage and analysis. They have worked with the Superfacility project to provide direct data reduction and analysis support to users at the microscope. Using a 100 Gbit fiber connecting the detector acquisition system and NERSC, they were able to reduce data reduction time by a factor of 2 (from 8 minutes to 4 minutes). This provides near-real-time feedback and also allowed them to free up local system resources to acquire more data than ever before. Further, they utilized Spin and other resources to provide a convenient web application[27] frontend to capture metadata and provide live feedback to the user at the microscope. The distiller app leverages NERSC Superfacility API to submit and monitor data reduction jobs on the real-time queue. NCEM's workflow is unique in that data is pulled directly from NCEM's data server into the compute allocation, which allows it to capitalize on fast job-local data storage solutions like Cori's DataWarp and *Perlmutter*'s all-flash scratch filesystem.

NCEM also uses Jupyter notebooks to provide interactive data analysis of the reduced data output. Users who previously needed to be familiar with ssh and command line tools are now able to process their data in real time using common workflows deployed on HPC infrastructure. The system can also be used during post-processing.

---

[27] https://distiller.lbl.gov

**Future plans**
The data reduction step can be further improved using *Perlmutter* for GPU computation. NCEM plans to utilize *Perlmutter*'s GPUs once testing of data retrieval and reduction has been fully completed. A real-time queue for this system would be absolutely necessary. NCEM also plans to implement full workflows from data generation to final output once suitable workflows have been identified. Another goal is to deploy the same workflow for other large data-generation systems at NCEM to better incorporate live processing of 10-100 GB datasets using image processing and AI/ML. A new detector with ~10x large data rate is planned to be installed in the coming years, requiring even more computation. Finally, NCEM plans to implement an automated data acquisition system that can acquire terabytes of data autonomously with live processing done at NERSC. The Superfacility project is essential to their future plans and workflows in order to deal with exponentially increasing data generation.

# 5: Lessons learned

One of the most important outcomes of the Superfacility project is that we have demonstrated unequivocally how the Superfacility model can benefit both science and ASCR facilities. This model of connected facilities is a growing part of the computing needs of SC user facilities and a driver for [new ASCR initiatives](#) such as the Integrated Research Infrastructure (IRI) Task Force.[28] In our requirements gathering, we found users increasingly want to integrate data analysis from experiments with simulations and AI. Far from being in competition with simulation and modeling, the Superfacility project has laid the groundwork for running complex workflows at an HPC facility, which will benefit a broad array of NERSC users.

Experiences and lessons learned from the Superfacility project have also changed the culture and perspective at LBNL. Superfacility use cases and drivers were core to the NERSC-10 Project Mission Need, and following the success of the Superfacility project, more science teams are engaging with ESnet and NERSC for their computing and networking needs. The Superfacility project can serve as a model for future work across SC, as we work toward a more connected infrastructure.

**Applying Team Science to understand requirements and guide ongoing technical work produced remarkable results.**
This project exemplified the spirit of team science, emphasizing collaboration and cross-pollination of ideas from disparate groups to produce generalized, reusable solutions. Combining the science and technical work areas was very successful in identifying cross-cutting requirements and sometimes unexpected synergies between work areas. The multiple rounds of explicit coordination between the technical work areas (e.g., running through work plans and upcoming milestones) helped align the work plans. For pre-existing projects, coming into this broader Superfacility project structure helped focus the development and deployment of tools. Having direct access to a portfolio of use cases and requirements was helpful, particularly as the project coordinated iterations with engaged science teams to improve tools.

---

[28] See "A Vision for ASCR Facilities" at https://science.osti.gov/ascr/ascac/Meetings/202109

The impact of this approach was visible in the series of Superfacility Demos we organized in 2020. By bringing together different science teams to demonstrate how they were using Superfacility-developed tools and the impact it was having on their science, we were able to inspire both the technical project team and the wider scientific community and gain momentum for the project goals.

**Visible management support is vital to the success of a cross-discipline project.**
We note the importance of management support for this project. The Computing Sciences Associate Lab Director, as well as ESnet and NERSC leadership, gave explicit and sustained support for this project, which gave the project leverage to keep this work prioritized across groups. This was particularly important during the COVID-19 pandemic; highlighting institutional priorities kept us focused.

**Dedicated liaisons for science teams kept our communications active and our developments on-target.**
One of the strongest aspects of the project was our science engagements. A single dedicated liaison between the science team and CSA staff meant that we could maintain a presence with the science teams rather than isolated and infrequent conversations. Having multiple areas of input kept our solutions responsive to the needs of each project, but also generalized to the wider science landscape, making our work more impactful and the results more sustainable. We intend to continue this close engagement with our science teams. As NERSC moves into the NERSC-10 project, it will remain deeply important for us to understand our users' needs and to maintain the close connection we have with the Superfacility teams.

**We need to work with the science teams' schedule and priorities, not impose our own.**
We noted that science teams sometimes did not have the resources to retool their processes and workflows to make use of Superfacility developments. Our science liaisons were able to help guide and prioritize their work, but progress was often linked to serendipitous external factors, such as a required software enhancement or hardware upgrade that presented an opportunity to make other improvements. This is a key point for future facility-driven work, realized early in the NESAP program: science teams do not necessarily have the time or personnel to take advantage of tools and technologies offered by facilities, even if that would make their workflows run more quickly and easily. Future projects should consider having dedicated funding to support the science teams to make changes to their workflows. In the future we also intend to consider hackathons as part of our engagement process – e.g., a half-day session with a science team as we work with them to deploy their tool.

**A flexible approach to project management is key.**

The nature of innovative R&D is that milestones are hard to define and hard to meet on schedule. Lots of unknowns in the technical work areas made projections difficult, especially initially, when work was first planned. In the first six months, many parts of the project were not meeting their first milestones, which was discouraging. We gradually got more comfortable with changing our milestones and adapting deliverables, and our flexible approach to project management was essential to meet this challenge. Our project management practice was somewhere between classical and agile; we stayed flexible enough to allow for changing requirements, delayed schedules, and sometimes whole work areas that became obsolete (for example, the scheduling simulator was abandoned once it became clear it was untenable). Project teams have noted that they benefited from the notion of structure, including the encompassment of all work areas under a comprehensive umbrella, even if that structure was somewhat fluid.

**It is hard to define clear milestones for science teams.**
For the science teams, it was harder to define "milestones." We aimed for a mix of technical demonstrations of use of a tool, and science achievements that could be made with the Superfacility toolset. These milestones were less distinct and harder to meet; we had to be very careful to define success in such a way that was useful, as well as achievable. They often had to be changed or updated as the needs of the science team changed. We have no influence over the speed of work or priorities of the science teams, but by continuously keeping them aware of new capabilities and demonstrating their benefits, we were able to help them take advantage of our newest developments.

**A complex project requires sophisticated, flexible management tools.**
The project management structure we chose was somewhat inadequate to the complexity and dynamism of the work we were doing. For example, it was hard for the technical work areas to track dependencies and see where their work connected to other areas as requirements and timelines shifted. Management tools designed to accommodate this continuous change could make this easier. We tracked schedules and changes in a simple set of spreadsheets; future projects should consider Gitlab, Asana, or other modern project management tools instead. It could also be hard for someone working in the project to see how all the pieces of the Superfacility project fit together, particularly in the beginning as we started to define our work. We would have benefited from a more centralized documentation system so people could more easily track progress in other work areas.

**A distributed, flexible project adapted well to a remote work environment.**
The distributed nature of our project lent itself to our new work-from-home environment, in the era of COVID-19. Our science teams are nearly all located remotely, so we were well situated to keep the engagements running smoothly when we had to switch to remote work. We learned to be flexible in our time frames. With an interconnected project, some pieces could be subject to delay for external reasons, and we learned to be generous with our expectations of other people's work plans.

**Long-term support requires explicit planning.**

The tools and technologies we developed in this project are intended to be useful far after the end of the project. We considered the longevity of this work from the very beginning and built it into our design principles of sustainable support. We aimed to reduce the amount of custom tooling done by project staff to avoid obsolescence if a staff member left. For example, in the Federated ID work we based the application on existing, industry-standard toolsets that would be externally maintained long-term. We also explicitly planned for long-term support as we reached the end of the project, with each work area completing a milestone to identify and train multiple people to support and maintain our technical work.

# 6: Future directions for Superfacility work

The Superfacility model of connected facilities is increasingly important for all DOE science. The Superfacility project was a vital first step in kick-starting and coordinating the work that is needed to support this model. Superfacility scientists now have a good toolbox of new tools, technologies and policies for developing their workflows at NERSC, but much work remains to be done. Superfacility-related work continues to be coordinated via a working group. The charge of this group is similar to that of the project: to inspire, coordinate, and communicate Superfacility work across science teams and technical work areas. It will maintain and refresh the portfolio of science engagements, serve as the coordination hub for tracking ongoing and future work, and spin up new work areas as new requirements are identified. A particular focus will be supporting our science engagements as they port their workflows to the NERSC *Perlmutter* system. We aim to demonstrate at least three Superfacility end-to-end workflows using *Perlmutter* in 2022, highlighting the key features:
- GPUs for data analysis
- Slingshot interconnect to bring data from an external facility in to the compute node
- SSD-based scratch filesystem for high IOPS (input/output operations per second) workloads.

## 6.1: Superfacility-related work for the NERSC-10 project

The NERSC-10 system project received CD-0 in late 2021 and aims to launch the next NERSC system in 2025. Superfacility science is a key driving use case for this system, and much of the lessons learned and technology developed in the Superfacility project will inform the design of NERSC-10. The Superfacility working group will help bootstrap NERSC-10's End-to-End workflows effort. Areas where on-going Superfacility work will directly impact the NERSC-10 project include:
- Deep engagements with science teams to define requirements and set benchmarks for workflows running on NERSC-10.
- Provide use cases that require incoming near-real-time requests for resources. A more dynamic NERSC-10 system may be needed to handle urgent data-processing needs.

- Drive explorations of object store hardware and software. The features of object stores (fast, responsive reads and writes for non-contiguous data) are very well suited to Superfacility science use cases, which typically have large, distributed datasets.
- Drive requirements for a closer linkage of the main compute system and "edge" or peripheral services that are currently supported by the Spin platform. Superfacility-type workflows require many services that will need to spin up and spin down, or remain persistent, depending on the needs of the experiment.

## 6.2: New Superfacility work

As noted throughout this report, although we have achieved a huge amount with the Superfacility project, many of the work areas still have challenges that will be addressed with future work. In this section we add to that list entirely new areas of work that would benefit Superfacility use cases.

**Engaging with new modes of science: distributed sensor networks.**
The Superfacility project would be almost impossible without the strong engagements we have with science teams. To maintain current and relevant requirements for future work, the Superfacility working group will continuously refresh the portfolio of science teams it engages with. An emerging technology area that will challenge the way HPC centers support experimental science is distributed sensor networks. These may be used (for example) to monitor environmental conditions at [a specific watershed](https://watershed.lbl.gov/).[29] The workflow for QA and analysis is complex; many different kinds of sensors may be involved, each with different needs for calibration and anomaly detection. Much computing may need to be done on low-power edge devices, and these need to be integrated with HPC centers running simulations, or training AI algorithms to be deployed on edge devices. For example, a simulation may guide where to place sensors in the field in quasi-real time to minimize model uncertainty.

**A mechanism for facility users to access NERSC without holding a full NERSC account.**
Superfacility science teams tend to have large collaborations, or are user facilities with their own userbases that number in the thousands. Setting up and managing full NERSC accounts is a burden on science teams and NERSC staff and will be hard to scale as the number of scientists turn to HPC. Many of those users need to access NERSC resources but do not need ssh-level access; it may be possible to support this with a limited account type that has less overhead for facility staff. For example, a user at the ALS needs to use NERSC to view and analyze data during their allocated beamline time via a standard analysis package, but does not need to edit any code. Work will be needed to define these new roles, and carefully account for the security implications.

**Capability for end-to-end troubleshooting for cross-facility workflows.**
Currently it is impossible to find out where a bottleneck exists in a cross-facility pipeline – whether it's a problem with the LAN, WAN, or data center. Superfacility teams would benefit enormously from a diagnostic tool for end-to-end troubleshooting these workflows. Work has already started on this topic, and a pilot demonstration is expected in late 2022.

---

[29] https://watershed.lbl.gov/

**Resilience improvements.**
Many superfacility science teams have experiments that operate 24/7 and need to have a plan in place for how to analyze their data when NERSC systems are unavailable. Despite the huge improvements made in user workflows resilience and facility and system resilience over the past few years, there is still room to improve. For example, although the NERSC-9 system supports rolling updates to compute node software, updates to the network can require small amounts of system downtime. A mitigation for this is to have a small section of *Perlmutter* that is capable of operating independently and can remain operational during all maintenance events, even during power outages. This is currently being considered at NERSC and will be a focus of effort for the NERSC-10 project.

Workflow resilience involves a number of factors; a workflow needs to be able to adapt when a system component becomes unavailable. API calls that provide system status information are useful for this, and can remove the need for a human in the loop. Ultimately, the best way to design a resilient workflow is to design a workflow that can run at multiple sites; when one is unavailable, it can simply switch to a new location and continue analysis. This is currently almost impossible, and much work is still needed to identify and mitigate the pain points in cross-facility workflows.

**Composable workflows.**
As the scientific community using HPC continues to expand, there is a risk that each science team will reinvent the wheel – creating their own custom data transfer tools, workflow orchestration managers, etc. This is hard for facilities to support and hard for science teams to develop and maintain. A true set of composable workflow components – or a set of standards that workflow components can adhere to – would be of real benefit. Some of this is being addressed in the ECP ExaWorks project, but building truly composable and interoperable workflows still needs a significant amount of community development and engagement.

**How to optimize an end-to-end workflow?**
The supercomputing world has well-established methods and guidance for optimizing single applications. It is much harder to optimize a full workflow that may include data transfer, several stages of simulations and/or analysis, and user interaction. For example, the NERSC-10 system aims to accelerate workflows and will have to grapple with how to define success in this area. Work is needed to set parameters and establish guidelines for how to profile workflows and how to focus effort on optimizing them.

## 6.3: Connecting Superfacility work to other ASCR facilities

There is an urgent need to connect workflows across all ASCR facilities – and indeed, to enable workflows to run across any available computing center. Achieving this goal is of course beyond the capabilities of any LBNL-focused work, but we have heard repeatedly from our science partners that they need to run their workflows at multiple computing sites. They need to run where the resources are the best fit for their needs, they need to burst to HPC when necessary, and they need to be resilient to any possible outages and maintenance periods. This includes local lab or university clusters, ASCR facilities, and the cloud. Several Superfacility teams are already working with cross-facility workflows and urgently need a more coordinated approach to computing across DOE computing centers. For example, JGI is developing a framework to execute their work at NERSC, LBNL IT, and EMSL, and LZ runs their analysis at NERSC and at their UK data center.

The Superfacility project has started to look into the practicalities required for this work, via the LDRD and ALCC-supported work described in Section 3.8. We have already identified several practical pain points (e.g., around cross-facility databases and access controls), which will be a useful starting point for future development work.

The importance of cross-facility workflows has been recognized by DOE SC, and an activity is planned for 2022 to design an architectural blueprint to develop these capabilities as part of the Integrated Research Infrastructure Task Force. LBNL will play a significant role in this process, and the hard work done by the Superfacility team and our science engagements will help guide the development of future DOE computing infrastructure.

# Acknowledgements

The authors would like to thank the science teams that worked closely with the Superfacility team throughout the project. Without their engagement, enthusiasm, feedback, and support, our work would not have been possible. In particular, we would like to thank Stephen Bailey, Stephen Chan, Jim Chiang, Jong Choi, Peter Ercius, Peter Denes, Kjiersten Fagnan, Heather

Kelly, Ralph Kube, Alex Hexemer, Maria Elena Monzani, Dula Parkinson, Amedeo Perazzo, Quentin Riffard, and Jana Thayer.

# Appendix A: Project management

The Superfacility project has been a unique undertaking at LBNL. It encompasses researchers from across divisions and directorates, working toward a common goal of enabling new discoveries by coupling experimental science with large-scale data analysis and simulations. Coordinating work across such a large group, and ensuring that the results would be coherent and useful to the science teams, took careful consideration and planning. In this section we discuss how we managed the Superfacility project, including the project structure, and how we handled our interactions with the wider group of stakeholders.

## A.1: Project principles

The Superfacillity project was designed to coordinate work being performed in a piecemeal fashion across different groups at LBNL. To avoid multiple one-off, specialized "demonstration" projects, and to ensure the Superfacility model could scale to all science teams who needed it, we needed to introduce this project structure around the existing work. By drawing requirements from multiple science teams we could ensure our work would be widely useful. We recognize that an increasing number of science teams need to operate a complex workflow at HPC facilities, but do not necessarily have the personnel to design a pipeline from scratch themselves. Equally, at facilities like ESnet and NERSC, we do not have the personnel to support multiple highly specialized workflows that each use custom tooling. We carefully planned our work in this project to meet those needs and produce generally useful tools and policies that could be supported in the long term. To this end, we planned our work along three principles:
- **Integration** of our work across multiple research and facility teams at LBNL
- **Scalability** of our work to multiple science teams – which also means scalable user support for an increasing number of users
- **Sustainability** of our work in the long term, by using industry standard tools where possible and planning to operationalise our work after the project ends.

## A.2: Project leadership

The Superfacility project leadership ensured that the team was focused and moving toward the successful completion of the project. The project was led by the following NERSC staff:

| Name | Role |
|---|---|
| Debbie Bard | Project Manager & L2 Lead for Applications: Requirements and Deployment |
| Cory Snavely | Deputy Project Manager & L2 Lead for Scheduling and Middleware |
| Jason Lee | L2 Lead for Automation and Networking |
| Lisa Gerhardt | L2 Lead for Data Management |
| Becci Totzke | Project Coordinator |

### A.2.1: Communication

Effective communication was the Superfacility project's primary tool for promoting cooperation, participation, coordination, and understanding between all stakeholders. Essentially every individual within NERSC, the CS Division, LBNL, the selected ASCR Facilities, and the selected Partnerships were stakeholders within the Superfacility project and were strongly interested in the project's positive outcome.

Communication throughout the project included the following formal meetings:

- Weekly management planning meetings
- Bi-weekly team meetings
- Monthly group leader updates
- LBNL leadership quarterly meetings
- Quarterly Executive Steering Committee meetings for the CS Area

### A.2.2: Schedule & Work Breakdown Structure

The Superfacility project used a combination of traditional waterfall and agile methodologies to manage the project schedule. This hybrid approach was successful in light of the unforeseen COVID-19 pandemic as it allowed the team to move milestones as needed due to the delays the science facilities were facing. Schedule status was updated monthly by the team and presented monthly at the Team meetings. See Appendix B for the Superfacility science goals at the end of 2021.

The project Work Breakdown Structure (WBS) Organizational Chart is shown in Figure 23. The Level 3 WBS dictionary is included in Appendix C.

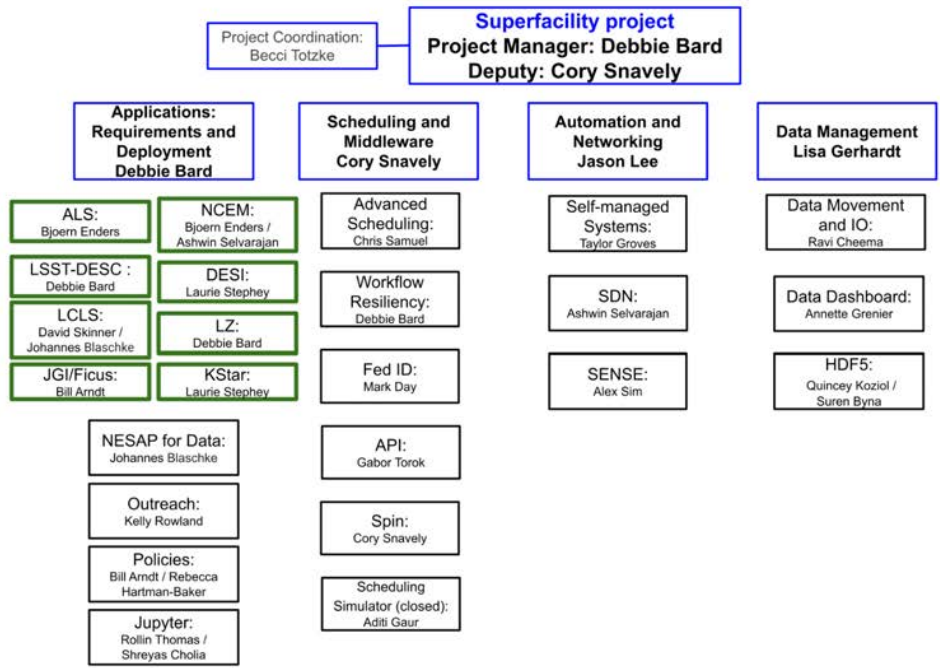

*Figure 23. Superfacility project WBS Organizational Chart.*

### A.2.3: Executive Steering Committee

The Superfacility project team is led by the Executive Steering Committee and includes other members from the Computing Sciences Area (CSA) at LBNL who are critical to the success of the project. The Executive Steering Committee worked in partnership with project members and others in the pursuit of Superfacility project goals. The project presented updates to the Committee once per quarter. This close communication with management stakeholders ensured that the project remained aligned with the CSA research priorities and goals and that we maintained strong support from our sponsors (particularly important in a project without specific funding). Committee members include:

- Executive Steering Committee
    - Jonathan Carter (LBNL), ALD for Computing Sciences
    - Sudip Dosanjh (LBNL), NERSC Division Director
    - David Brown (LBNL), CRD Division Director
    - Inder Monga (LBNL), ESnet Division Director
    - Katie Antypas (LBNL), NERSC Division Deputy and Data Department Head
    - Peter Nugent (LBNL), CRD Deputy for Scientific Engagement
    - Chin Guok (LBNL), ESnet Network Planning Team Lead

## A.3: Encompassing ongoing and related work

When the Superfacility project was initiated as a result of the CS Area Strategic Plan in 2018, it was clearly acknowledged that a lot of work was already happening in the space and had been ongoing for several years. A key aim of the project was to act as a communication channel for this existing work, to broaden its impact, and to ensure work was not being inadvertently replicated across groups. An example of this is the work to support and add features to Jupyter services at NERSC (see Section 3.4). The Usable Software Systems group in the Scientific Data Division at Berkeley Lab has been collaborating with NERSC and with the Jupyter development team for several years to create new features relevant to science partners (e.g., NCEM). What the Superfacility project could offer to this on going work was additional detailed science use cases so that the development of new features (e.g., task parallelization with Dask) could be generalized and tested by multiple teams of scientists.

Not all Superfacility-related work (as identified in the CS Area Strategic Plan) was brought into the project. We needed to keep the project focused and to have a goal that would be achievable within the three years of the project. However, we also needed to keep up to date with other related work at LBNL (and more widely) and to connect with it where appropriate. The regular meetings with the Executive Steering Committee helped to keep us current with other work. This became particularly important during the Covid-19 pandemic, when casual hallway conversations (which would usually connect us to other researchers) became harder to arrange. For example, we added the KSTAR team mid-way through the project as it became clear that their emerging use case was important and would test our capabilities in a way that was complementary to the initial set of science engagements.

### A.3.1: Superfacility and the NERSC-9 project

An important enabler of Superfacility-related work is the NERSC-9 project, which opened the *Perlmutter* supercomputer to NERSC users during 2021. One of the key project goals for our third year (2021) was to get our science teams up and running on Perlmutter as soon as possible, by deploying our tools on the new system and ensuring that our science teams had the appropriate access and system capabilities they needed. We carefully tracked certain items in the NERSC-9 project that were of particular importance to Superfacility teams, including:
- API deployment
- Spin connections to *Perlmutter*
- Globus
- CVMFS
- Scratch file system performance
- Scalable AI and analytics platforms, including Jupyter
- Queue configuration for near-real-time access

Due to the delayed delivery of the *Perlmutter* system, we were not able to demonstrate all Superfacility teams' end-to-end workflows running on the full system by the end of the project. This goal is a high priority for NERSC and for the NERSC-9 project, so we will continue to track and coordinate this work after the end of the project with a subset of the project team. This additional coordination will end with the acceptance of NERSC-9.

## A.4: Project support

A vital part of the success of the Superfacility project is that we had buy-in from all parts of LBNL. Our science partners and sponsors at the LBNL management level understood the importance of coordinating this work to increase its impact and supported researchers and engineers spending their time working on it. LBNL (and in particular the CS Area, under the leadership of Kathy Yelick) has been a champion of the Superfacility model for several years, so the urgency of this work was already well established. We did not need to convince leadership that this was worthwhile doing. The benefits of coordinating the existing work were clear.

Staff efforts on the Superfacility Project were supported by the NERSC and ESnet programs. It was a key strategic priority for NERSC and ESnet, so most of our work was included in base funding profile. We had no additional budget or grants to support our work. The work was mainly undertaken by facility staff at NERSC and ESnet as part of our mission to support science and by researchers who had existing funding to work on these topics. Additionally, some of the resilience work was supported by an LDRD grant, awarded in the second year of the project. Given this, the visible support of our leadership was essential to driving the work forward.

Another important way we built support for the project was through our ongoing relationships with our science engagement teams. We didn't simply ask them what they needed and then deliver what we thought they said; we had a long series of conversations and iterations over the whole course of the project to produce something that would be genuinely useful to them. This was a lengthy trust-building exercise that has resulted in a much stronger infrastructure for science. Building these relationships is already benefiting new NERSC projects; for example, we have enthusiastic partners for our NERSC-10 project (e.g., to provide detailed requirements and prototype benchmarking codes). Our science partners have also learned more about NERSC and ESnet and what we can offer, making them more effective in doing their science at ASCR facilities.

# Appendix B: Superfacility Science Goals

| WBS | Task Name | Start | Finish |
|---|---|---|---|
| | **Superfacility project** | | |
| **1.00.01** | **Tier 1 Milestones - Overall Project Goals** | | |
| 1.00.01.01 | By the end of CY 2021, 3 (or more) of our 7 science application engagements will demonstrate automated pipelines that analyze data from remote facilities at large scale, without routine human intervention, using these capabilities:<br>- Real-time computing support<br>- Dynamic, high-performance networking<br>- Data management and movement tools<br>- API-driven automation<br>- Authentication using Federated Identity | | |
| **1.00.02** | **Tier 2 Milestones - Science Goals: Every science area should have 3 demonstrations total (at least one/year)** | | |
| 1.00.02.01 | ALS | | |
| ALS.1 | #1: Demonstrate high throughput data transfer and restrictive sharing via globus endpoint | **05/06/19** | **05/06/19** |
| ALS.2 | #2: Successfully demonstrate Live Processing and Feedback for Small Ptychography Run | **03/30/20** | **03/30/21** |
| ALS.3 | #3: New science gateway launched to make data accessible using FedID for selected projects on at least one beamline | **06/30/20** | **09/30/21** |
| ALS.4 | #4: Demonstrate automatic archiving | **12/31/21** | **12/31/21** |
| ALS.5 | #5: Synchronization of schedules or pipeline resiliency | **12/31/21** | **12/31/21** |
| ALS.6 | #6: Science gateway: Instruments Data sharing with ALS databases | **05/01/10** | **12/30/20** |
| 1.00.02.02 | LSST-DESC (Broker for transient feeds, catalog analysis) | | |
| LSST.1 | #1: Decision point about what level of broker will be hosted at NERSC, based on technical capabilities available in Spin, SDN, Data Management and Policy | **06/01/19** | **06/01/19** |
| LSST.2 | #2: Demonstrate hundreds of simultaneous users running Jupyter notebooks to analyze DC2 data from DESC (done on July 18) | **07/20/19** | **07/20/19** |
| LSST.4 | #4: Show managed data from DC2 simulation campaign with data transfer tools and sharing with globus sharing | 07/31/20 | 07/31/20 |
| LSST.5 | #5: At-scale testing of the image processing workflow, coordinated via Spin | 07/01/21 | 12/31/21 |
| LSST.6 | #6: Define requirements for setting up Rubin Science Platform at NERSC | 07/01/21 | 12/31/21 |
| LSST.7 | #7: Make plan for large data transfers from SLAC | 07/01/21 | 12/31/21 |
| LSST.8 | #8: Long-term support: Identify a backup DESC support person and make a plan for training them | 11/01/21 | 12/31/21 |
| 1.00.02.03 | LCLS (X-ray FEL images living bacteria, chemical bonds. Complex experiments with many users who benefit from near real-time feedback) | | |
| LCLS.1 | #1: ECP SDN milestone (transfer using two paths) and using SENSE (jason will find the link between SDN and SENSE) | 01/11/19 | 01/11/19 |
| LCLS.2 | #2: Demo of combined SDN automation and adaptive Cori scheduling for first LCLS-I runs | 08/15/19 | 08/31/21 |
| LCLS.4 | #4: Provision bandwidth beyond ESnet border | 11/21/21 | 11/21/21 |
| 1.00.02.04 | JGI (Support user facility operations for sequencing and genome assembly) | | |
| JGI.1 | #1: Transfer all possible JGI services to Spin | 01/01/19 | 07/26/19 |
| JGI.2 | #2: Use Globus sharing for JAWS workflow management | 01/01/19 | 03/30/20 |
| JGI.3 | #3: Use Federated ID to log into NERSC | 01/01/21 | 12/31/21 |
| JGI.4 | #4: Test GHI for automated data movement to HPSS | 07/31/19 | 01/02/20 |
| 1.00.02.05 | NCEM | | |
| NCEM.1 | #1: NERSC deploys high-performance network / SDN capabilities for NCEM | 12/20/18 | 12/20/18 |
| NCEM.2 | #2: NCEM 2019 Summer Run with Clean performance | 07/31/19 | 07/31/19 |
| NCEM.3 | #3: Demonstrate near-real-time scheduling on experiments using 4D-STEM | 09/30/20 | 09/30/20 |
| NCEM.4 | #4: Enable data movement for experiments: Fast Feedback and Data Archiving | 02/14/22 | 02/14/22 |
| NCEM.5 | #5: Resilience planning when NERSC is down | 08/31/20 | 08/31/20 |
| NCEM.6 | #6: Analyze 4D-STEM experimental data with Jupyter | 03/15/20 | 03/15/20 |
| NCEM.7 | #7: Rewire NCEM PCs for fast connection to NERSC, demo reduction pipeline | 06/01/21 | 10/25/21 |
| 1.00.02.06 | DESI | | |
| DESI.1 | #1: Tested QA/quicklook framework going in Spin at SpinUp Workshop #2 | **09/17/18** | **09/17/18** |
| DESI.2 | #2: Accelerate the DESI extraction code so it is feasible to use KNL | **01/01/19** | **01/01/19** |
| DESI.6 | #6: Ability to resize jobs (remove nodes during the job) | **09/30/20** | **09/30/20** |
| DESI.7 | #7: Ability to use globus as collaboration account | **09/30/20** | **09/30/20** |
| DESI.8 | #8: Be able to efficiently transfer large amounts of data (both in volume and number of files) between file systems while retaining file ownership, permissions, and timestamp metadata | **04/30/21** | **05/28/21** |
| DESI.9 | #9: NERSC to provide some kind of nightly, incremental backup of important directories | **12/31/21** | **12/31/21** |
| **1.00.02.07** | **LZ** | | |
| LZ.1 | #1: Resiliency plan for data archiving | **07/09/19** | **07/09/19** |
| LZ.2 | #2: Run MDC3 re-processing, mimicking the real-time needs of the experiment. Will test: real-time processing, workflow failover (for both data movement and compute), reservations. | **12/31/19** | **12/31/19** |
| LZ.3 | #3: Run MDC3 production | **12/31/19** | **12/31/19** |
| LZ.4 | #4: Profile and accelerate the LZ analysis code for use on GPUs | **04/01/21** | **04/01/21** |
| LZ.5 | #5: Real-time operations | **05/01/21** | **12/31/21** |
| LZ.6 | #6 Demonstrate workflow failover for both data movement and compute | **04/01/21** | **07/12/21** |
| LZ.7 | #7 NESAP project: determine plan for future development work | **05/01/21** | **11/30/21** |
| LZ.8 | #8 Long-term support: identify backup person to hold periodic check ins | **05/01/21** | **12/31/21** |
| **1.00.02.08** | **Fusion PPPL/ORNL/KSTAR** | | |
| KSTAR.3 | #3: Improve performance of data analytics on GPUs | **01/01/21** | **09/01/21** |
| KSTAR.4 | #4: Create improved Spin/Jupyter dashboard to visualize results | **01/01/21** | **09/01/21** |
| KSTAR.5 | #5: Implement shifter on DTNs | **05/20/21** | **12/31/21** |
| KSTAR.6 | #6: Switch to minIO to manage data | **05/20/21** | **12/31/21** |

*Items that are highlighted green are complete.

# Appendix C: WBS Dictionary

| Level | WBS Code | WBS Name | Goal | Description | Lead |
|---|---|---|---|---|---|
| 1 | 1.00.01 | Superfacility project (11/22/21) | By the end of 2021, at least 3 of the 7 science use cases are analyzing data between at least 2 facilities in an automated fashion and at production level scale.<br><br>Demonstrate automated pipelines that analyze data from remote facilities at large scale, without routine human intervention, using these capabilities:<br>- Real-time computing support<br>- Dynamic, high-performance networking<br>- Data management and movement tools<br>- API-driven automation<br>- Authentication using Federated Identity | Partner with three diverse experimental facilities, one ASCR compute center, and ESnet to demonstrate production end-to-end automated pipelines at NERSC.<br><br>- Deploy large scale computing and storage resources<br>- Provide reusable building blocks for experimental scientists to build pipelines<br>- Provide scalable infrastructure to launch services<br>- Provide expertise on how to optimize pipelines<br><br>Initiative: We will implement the superfacility model, collaborating with Office of Science User Facilities to enable seamless and high performing end-to-end pipelines. | D. Bard / C. Snavely |
|  | 1.00.02 |  | Science Goals (Level 2 Milestones) | Every Science area should have milestones with approximately 3 demonstrations total (at least one/year).<br><br>Partner with three diverse experimental facilities, one ASCR compute center, and ESnet to demonstrate production end-to-end automated pipelines at NERSC. | D. Bard |
|  | 1.00.02.01 | ALS | ALS (Advanced Light Source) is an x-ray synchrotron facility used to image a variety of things. | Beamline with ephemeral users who need near-real-time feedback and scalable analytics. Data needs to be transferred to NERSC, analyzed, backed up in the archive, and be shared with a limited set of users after processing is finished. | B. Enders |
|  | 1.00.02.02 | LSST-DESC | Create a broker for next-day consumption of transient feeds and catalog analysis using the stream of LSST telescope imaging data | Take imaging data, pick out interesting parts by querying an external database, and form channels that will be consumed to search for transient events like supernovae. The broker program is still being formed, expected to be fully online by 2021. | D. Bard |
|  | 1.00.02.03 | LCLS | X-ray FEL (free electron laser) images living bacteria, chemical bonds, etc. Complex experiments with many users who benefit from near real-time feedback | Beamline with ephemeral users who need near-real-time feedback and scalable analytics for large data volumes. | D. Skinner / J. Blaschke |
| | 1.00.02.04 | *JGI / Ficus / BioEPICS* | Characterize metagenome populations across environments, detail of comparison scales with CPU resources | The majority of JGI science portals are now hosted in SPIN and routinely benefit from it's proximity to NERSC storage and compute resources. The largest and most difficult Exabiome metagenome datasets are now being processed at NERSC using resource reservations to schedule. The JGI Jaws workflow management tool relies on Superfacility Globus features to migrate data and processing tasks between different DOE compute facilities. | B. Arndt |
| | 1.00.02.05 | *NCEM* | Development of a Fast Framing Detector for Electron Microscopy | NCEM will start production runs in the summer / fall of 2019. They will need a near real-time feedback process to allow them to 'steer' the experiment. | A. Selvarajan |
| | 1.00.02.06 | *DESI* | DESI (Dark Energy Spectroscopic Instrument) will measure the effect of dark energy on the expansion of the universe. Needs data management tools and deadline driving scheduling. | DESI will start streaming data in 2019. Complex pipeline triggered by the arrival of data on the system. Data is owned by a collaboration user, so needs to transfer data in and maintain collab ownership. Needs deadline driving scheduling to assess quality of the previous night's observations in order to decide on plan for the next night (i.e. they need all the results by mid day, but not immediately as they come in) | L. Stephey |
| | 1.00.02.07 | *LZ* | LZ (Lux Zepplin) dark matter experiment turns on in June 2020. It will need real-time processing to measure detector health, data management tools, and advance scheduling with resilience failover. | Search for dark matter using a direct detection experiment 1 mile underground in South Dakota. Data taking starts 2021, and the LZ team will also run regular simulation campaigns. LZ will use NERSC for rapid data analysis to monitor data quality, using dashboards on Spin and the real-time queue. | D. Bard |
| | 1.00.02.08 | *KStar* | KSTAR is a fusion energy experiment (tokamak) in Korea. American collaborators from PPPL and ORNL are working to develop the Delta framework which will analyze data from KSTAR in near real time, between tokamak discharges, using HPC resources like NERSC. | KSTAR is a working experiment that usually runs in "campaigns" that last several months. KSTAR has some large datasets like their turbulence diagnostics that are too large to be analyzed between shots with local compute resources. Delta streams the data to Cori, performs CPU or GPU based processing, and uploads the results to an interactive Spin dashboard. On Perlmutter KSTAR will require open networking to allow data to be streamed from Korea. They rely on Shifter for their complex stack, so they also request that Shifter be installed on the DTNs. | L. Stephey |
| **2** | **1.01** | **Applications: Requirements and Deployment** | | | **D. Bard** |
| 3 | 1.01.01 | *NESAP for Data* | Help experimental facilities who have data-intensive computing problems get ready for Perlmutter | The NESAP program has been successful in preparing science teams for KNL architecture. In the next iteration of NESAP, teams will be supported in their transition to the GPU architecture of Perlmutter. In the NESAP for Data branch, experimental facilities are | J. Blaschke |

| | | | | particularly encouraged to apply for help in optimising their complex workflows. Successful proposals will be given access to hardware experts, training sessions and early access to Perlmutter, and proposals in the top tier will in addition be given a NESAP postdoc to work on their project. | |
|---|---|---|---|---|---|
| 3 | 1.01.02 | Outreach | Create and execute a communications plan to convey the work being done in the Superfacility project to three main audiences:<br>1) Users<br>2) DOE program managers<br>3) Other compute facilities | An essential element to the success of the Superfacility project is ensuring we are able to communicate our plans and the opportunities to our user community, and to the DOE program managers that have stewardship over the DOE experimental and observational facilities. We also have an opportunity to coordinate our work with the LCFs (and other lab-based computing groups). This kind of working relationship will be essential as many experiments will need to run at multiple compute sites, and we need to make it as seamless as possible for them to do so. | K. Rowland |
| 3 | 1.01.03 | Policies | Identify NERSC policies that need to be implemented, changed or replaced in order to support experimental science workflows at NERSC. | Policies cover many elements of work at NERSC, including who can get accounts and what access users have to NERSC resources. This sub-project will initially focus on user account policies, with the aim of laying out the policies changes required (if any) to support different ways of using NERSC, for example: access to a limited set of actions via an experiment account (rather than an individual user account); access to data only (no compute); and access to NERSC via Jupyter. Additional policy areas may come up as the work continues; this sub-project will evolve in scope accordingly. | B. Arndt / R. Hartman-Baker |
| 3 | 1.01.04 | Jupyter | Make Jupyter (Notebooks, JupyterLab, and JupyterHub) into a superfacility user interface. | Scientists love Jupyter because it combines documentation, visualization, data analytics, and code into a document they can share, modify, and even publish. Jupyter is becoming an essential interface to NERSC and other high-performance computing (HPC) centers. This sub-project defines what adaptations and enhancements are required to make Jupyter (Notebooks, JupyterLab, JupyterHub, etc) into a one-stop portal for experimental and observational science in the superfacility. | R. Thomas / S. Cholia |
| **2** | **1.02** | **Scheduling and Middleware** | | | **C. Snavely** |
| 3 | 1.02.01 | Scheduling Simulator | Develop a simulator for the NERSC workload manager that will enable us to better understand and quantify the impact on NERSC system productivity from changes related to supporting data-oriented workflows. Conditions to simulate | New workflows from experimental facilities have the potential to disrupt the full NERSC workload in unexpected ways. For example, regular incoming experimental data with short-turnaround analysis requirements could prevent full machine jobs from ever starting. In order to understand what kinds of workloads | A. Gaur |
| | | | include a) new job submission patterns and b) proposed scheduling policy changes. | NERSC can support, we need to study their impact on different classes of jobs. | |
| 3 | 1.02.02 | Advanced Scheduling | Integrate real-time and short-turnaround job scheduling into the NERSC workload manager while maintaining high overall utilization. Develop and integrate schedulable storage and bandwidth resources into the NERSC job scheduling. | Several experimental facilities require real-time or short-turnaround processing of data from the experiment. Supporting these workloads while maintaining high system utilization and minimizing the impacts on the NERSC workload will require innovations in SLURM and a careful definition of the queue parameters and QOS'. This group will also work on developing schedulable resources into the NERSC ecosystem beyond compute time, such as bandwidth to storage and network. | C. Samuel |
| 3 | 1.02.03 | Workflow Resiliency | Define NERSC policy WRT workflow resiliency, including mitigation strategies for NERSC downtimes, what tools to support to enable transferable workloads to other sites/services, and come up with a plan for how to communicate the NERSC outage schedules and our advice/policy to users. | One of the top requirements from experimental facilities is to increase NERSC uptime. Given that downtime (scheduled and unscheduled) is inevitable, this group will define what information should be communicated to users in the event of an outage (via the API, for example). It will also write and maintain advice for users on resiliency planning.<br><br>This group will also study the feasibility of various mitigation strategies, including:<br>- isolating a part of Cori/Gerty for use for real-time workloads during an outage;<br>- enabling workflows to run at other DoE computing sites<br>- enabling workflows to run on AWS | D. Bard |
| 3 | 1.02.04 | FedID | Leverage identity federation techniques to provide seamless access to data and services at NERSC using credentials from partner DOE facilities and/or a scientist's home institution. | (1) Users of a DOE experimental facility who also have a NERSC account, should be able to use their identity from the experimental facility or their home institution to authenticate to NERSC services, as well as obtain credentials that can be used to call the NERSC Superfacility API.<br><br>(2) Users of a DOE experimental facility who have been validated by that facility, but do not have a NERSC account, should be able to use the identity from their home institution or experimental facility to login to a facility-specific portal or gateway at NERSC to access their experimental data as well as run pre-selected compute jobs authorized by the experimental facility. | M. Day |

| 3 | 1.02.05 | *Superfacility API* | Deploy an API that allows experiments to automate their workflows with scripts or workflow managers rather than manual command-line methods. The API endpoint should support methods to access information about NERSC systems and perform tasks such as data transfer, job submission, etc. required for workflow operation. | Experimental facilities have complex multi-stage workflows that are often hard to control from their remote site, without access to the NERSC command line or involving several people in the loop. Examples of this include reserving compute time for an experimental run, finding out if NERSC compute or filesystems are down, … This group will identify the common tasks that our target experiment engagements need to perform, and develop an API that they can program against to perform these tasks.<br><br>Previous in first draft:<br>Develop an API to enable users to use automation to orchestrate workflows that use NERSC resources.<br><br>- NEWT: API for job submission, authentication, quota checking<br>- Globus transfer and optimized DTNs for easy transfer between facilities<br>- real-time queues to let jobs start immediately<br>- Queues to accommodate all job sizes<br>- Globus sharing test deployment to let users efficiently share data and control access<br>- Software Defined Networking for configurable networking<br>- Shifter: Docker for HPC allows users to deploy containers | G. Torok / J. Riney |
|---|---|---|---|---|---|
| 3 | 1.02.06 | *Spin* | Provide a container-based infrastructure that can be used a) by staff, to deploy services that complement NERSC compute and storage systems, and b) by users, to build and manage services that support their research projects. | Staff and users will be able to use a Docker-based infrastructure to deploy services that are performant and tightly integrated to NERSC compute and storage systems. Documentation and training are available. Implementation options exist for user-managed services, NERSC-managed user-facing services, and internal NERSC services. | C. Snavely |
| **2** | **1.03** | **Automation** | | | **J. Lee** |
| 3 | 1.03.01 | *Self-managed Systems* | Provide the **link** between the work being performed by a diverse group of people into many different kinds of self-driving systems and the Superfacility project. | The self-managed systems discussion group was established as an outcome of the CS Area Strategic initiative for self-driving systems. This group comprises people doing work in the area of monitoring systems, and making decisions based on the results of that monitoring. The topics under discussion in this group include IO (Glenn Lockwood), memory (Eric Roman), networking (Mariam Kiran, Taylor Groves) and systems monitoring (Tom Davis/Cary Whitney).<br><br>This sub-project will broker connections between this discussion group and the Superfacility project. For example, the discussion group will be looped into the requirements conversations | T. Groves |

| | | | | with the Superfacility experimental partners so that their use cases can motivate and guide the work being done for self-managed systems. And this sub-project will keep the Superfacility team informed about work being done in self-managed systems, flagging opportunities for value-added and collaboration to other Superfacility sub-projects. | |
|---|---|---|---|---|---|
| 3 | 1.03.02 | *Software Defined Network (SDN)* | Programmatically enable seamless transfer mechanisms for experimental facilities to stream data into Cori and Perlmutter. Software control of the network (SDN) should allow new flexibility in how to plan the logistics of their data, particularly in regard to bandwidth. The data transfer mechanisms scientists can include in their workflows are necessarily limited by the linkages between resources, schedulability of resources, and organizational policy. | Experimental workflows often require data to be streamed from an external source or detector, directly to a compute node where it can be analyzed in real time to give feedback to the experimenter. Bottlenecks can arise in the networking while sending data from the experiment to NERSC, and this group will develop SDN tools to enable easy and performant data flow into NERSC. An overriding, and perhaps novel, concern in SDN controls is making the "best use" of total bandwidth and metrics related to the policy around those controls. Contention for resources is a well-known concern in HPC. As it regards bandwidth as a resource similar principles of time and space sharing apply. Separation of performance concerns and reliable schedulability of resources through a batch queue scheduler are "old school", and should be addressable through software.

From a NERSC-mostly perspective it will become increasingly useful to offer the bandwidth into our HPC machines as a shared and allocated resource that science teams can plan around, make well-motivated proposals for, and expect to happen in the delivery of their science.

Enable experimental facilities to stream data directly to compute nodes during experimental running | A. Selvarajan |

|   | | | | | |
|---|---|---|---|---|---|
|   | *1.03.03* | *SENSE* | The Software-defined network for End-to-end Networked Science at Exascale (SENSE) project is building smart network services to accelerate scientific discovery in the era of 'big data' driven by Exascale, cloud computing, machine learning and AI. The goal is to dynamically build end-to-end virtual guaranteed networks across administrative domains, with no manual intervention. In addition, a highly intuitive "intent" based interface, as defined by the project, allows applications to express their high-level service requirements, and an intelligent, scalable model-based software orchestrator converts that intent into appropriate network services, configured across multiple types of devices. | Network designs are evolving at a rapid pace toward programmatic control, driven in large part by the application of software to networking concepts and technologies, and evolution of the network as a key subsystem in global scale systems, such as those serving major science collaborations that incorporate large scale distributed computing and storage subsystems. However, even the most optimistic projections of software adoption and deployment do not put networks on a path that would make them behave as a truly smart or intelligent system from the application or user perspective, nor one capable of interfacing effectively with facilities supporting highly automated workflows at sites located across the world. Today, domain science applications and workflow processes are forced to view the network as an opaque infrastructure into which they inject data and hope that it emerges at the destination with an acceptable Quality of Experience. There is little ability for applications to interact with the network to exchange information, negotiate performance parameters, discover expected performance metrics, or receive status/troubleshooting information in real time. As a result, domain science applications frequently suffer poor performance. It is clear that current static, non-interactive network infrastructures currently do not have a path forward to assist or accelerate domain science application innovations.<br><br>This group will work on building smart network services supporting interactions between the application and the network. | A. Sim |
| **2** | **1.04** | **Data Management** | *Enable seamless movement of data between storage layers.* |   | **L. Gerhardt** |
| 3 | *1.04.01* | *Data Movement Tools* | Enable seamless movement of data between storage tiers at NERSC | Experimental groups have workflows that require the movement of data from external sources to the NERSC community filesystem and to archive during the running of an experiment or stage to the fast storage tier for analysis. Experiments also require the use of archived data for analysis or sharing post-experiment. This group is developing tools that allow the easy movement of data between the storage layers at NERSC.<br><br>These tools will also integrate into the Superfacility API, allowing users to access and move their data at NERSC without using the command line. | R. Cheema |

| 3 | 1.04.02 | Data Dashboard | NERSC users will be able to view and manage their data at NERSC via a simple and beautiful web interface | Managing data at NERSC is difficult; data can be spread between multiple filesystems and across multiple projects. In particular, it can be hard for a PI to track usage of their group storage resources. The Data Dashboard is designed to provide a simple and instinctual view of a user's data at NERSC and provide an interface for common actions (chmod, chgrp, transfer, etc.) | A. Grenier |
|---|---|---|---|---|---|
| 3 | 1.04.03 | HDF5 | Support sparse data management natively in HDF5 | Working with sparse matrices and other sparsely-populated data structures is a common need in many science, engineering, and mathematical domains, but is frequently overlooked and underserved by data management frameworks and storage packages. HDF5 does not currently store sparse data in an efficient, performant, or portable way. Exploring efficient data representations and designs and developing efficient native support for sparse data storage to HDF5 is of importance for various use cases, including NCEM. This will enable reduced file sizes, faster I/O performance, full access to the entire HDF5 ecosystem of tools, and better adoption of sparse data storage in science communities where it's the best match to their data storage use cases. | S. Byna |